\documentclass[aps,twocolumn,english,preprintnumbers,amsmath,amssymb,nofootinbib,superscriptaddress]{revtex4-2}

\usepackage{hyperref}
\usepackage{graphicx}
\usepackage{xcolor}
\usepackage{bbold}
\usepackage{amsmath}

\newcommand{\beq}{\begin{equation}}
\newcommand{\eeq}{\end{equation}}
\newcommand{\bqa}{\begin{eqnarray}}
\newcommand{\eqa}{\end{eqnarray}}

\def\sumint{\hbox{$\sum$}\!\!\!\!\!\!\!\int}
\def\square{\vcenter{\vbox{\hrule height.4pt
          \hbox{\vrule width.4pt height4pt
          \kern4pt\vrule width.3pt}\hrule height.4pt}}}

\newcommand{\symff}{\text{SYM}_{4,4}}

\begin{document}

\title{ Scheme dependence of two-loop HTLpt-resummed $\symff$ thermodynamics}

\author{Qianqian Du}
\affiliation{Department of Physics, Guangxi Normal University, Guilin, 541004, China}
\affiliation{Guangxi Key Laboratory of Nuclear Physics and Technology, Guilin, 541004, China}
\affiliation{Institute of Particle Physics and Key Laboratory of Quark and Lepton Physics (MOS), Central China Normal University, Wuhan, 430079, China}

\author{Michael Strickland}

\author{Ubaid Tantary}
\affiliation{Department of Physics, Kent State University, Kent, OH 44242, United States}

\date{\today}

\begin{abstract}
The resummed thermodynamics of ${\cal N}=4$ supersymmetric Yang-Mills theory in four space-time dimensions ($\symff$) has been calculated previously to two loop order within hard thermal loop perturbation theory (HTLpt) using the canonical dimensional regularization (DRG) scheme. 
Herein, we revisit this calculation using the regularization by dimensional reduction (RDR) scheme.
Since the RDR scheme manifestly preserves supersymmetry it is the preferred scheme, however, it is important to assess if and by how much the resummed perturbative results depend on the regularization scheme used.
Comparing predictions for the scaled entropy obtained using the DRG and RDR schemes we find that for $\lambda \lesssim 6$ they are numerically very similar.  We then compare the results obtained in both schemes with the strict perturbative result, which is accurate up to order $\lambda^2$, and a generalized Pad\'{e} approximant constructed from the known large-$N_c$ weak- and strong-coupling expansions.  Comparing the strict perturbative expansion of the two-loop HTLpt result with the perturbative expansion to order $\lambda^2$, we find that both the DRG and RDR HTLpt calculations result in the same scheme-independent predictions for the coefficients at order $\lambda$, $\lambda^{3/2}$, and $\lambda^2 \log\lambda$, however, at order $\lambda^2$ there is a residual regularization scheme dependence.
\end{abstract}


\maketitle

\section{Introduction}

Supersymmetric field theories have generated a great deal of interest in the past decades \cite{Gervais:1971ji,Volkov:1973ix,Akulov:1974xz,Nilles:1983ge,Martin:1997ns}. Such field theories are invariant under supersymmetry transformations in which bosonic and fermionic degrees of freedom are transformed into one another.  Although there is currently no experimental evidence that such theories are realized in nature, they have recently proven to be useful due to the ability to employ the conjectured (and strongly evidenced) holographic duality between strongly-coupled conformal field theory (CFT) in the large-$N_c$ limit and weakly-coupled gravity in five-dimensional anti-de Sitter space (AdS) \cite{Maldacena:1997re}.  This conjectured mathematical equivalence is called the AdS/CFT correspondence and, in the context of thermodynamics, has been used to calculate the strong-coupling limit of ${\cal N}=4$ supersymmetric field theory in four space-time dimensions ($\symff$) \cite{Gubser:1998nz}.  In this paper, we focus on resummed perturbative calculations of thermodynamics in $\symff$ which complement such calculations.

One of the motivations for the calculation presented herein is to test methods that have been applied in the context of finite temperature and density quantum chromodynamics (QCD).  In particular, we would like to apply perturbative reorganizations which have been used to improve the convergence of the successive perturbative approximations to the QCD thermodynamic potential \cite{Andersen:1999fw,Blaizot:1999ip,Andersen:1999sf,Blaizot:1999ap,Andersen:1999va,Blaizot:2000fc,Peshier:2000hx,Blaizot:2001vr,Andersen:2002ey,Andersen:2003zk,Blaizot:2003iq,Andersen:2009tc,Andersen:2010ct,Andersen:2010wu,Andersen:2011sf,Andersen:2011ug,Haque:2012my,Haque:2013qta,Haque:2013sja,Haque:2014rua,Andersen:2015eoa,Haque:2020eyj}.  These studies have demonstrated that it is possible to obtain excellent agreement with continuum extrapolated lattice calculations of QCD thermodynamics for $T~\gtrsim~250-300$ MeV using such methods.  Unlike $\symff$, QCD is a confining theory at low temperature, however, at high temperature there are many similarities between $\symff$ and QCD.  This stems from the fact that (a) QCD is asymptotically free and (b) the two theories are similar in the weak coupling limit. Comparing perturbative $\symff$ and QCD one finds that (1) the forms of gluonic and fermionic collective modes are the same, and the scalar collective modes are those of a massive relativistic particle; (2) the transport coefficients, such as the shear viscosity $\eta$, which are dominated by the Coulomb-like interactions, are quite similar~\cite{Huot:2006ys}; and (3) the energy loss and momentum broadening of highly energetic test particles are also alike~\cite{Czajka:2013lla}. These studies indicate that the key difference between perturbative finite-temperature QCD and $\symff$ is the number and type of degrees of freedom, with $\symff$ having, in addition to the adjoint gauge field, four adjoint Majorana fermions and six adjoint scalars.

Unlike QCD, however, $\symff$ is ultraviolet finite due to its supersymmetric nature and has a vanishing $\beta$ function.  Due to this, the `t Hooft coupling $\lambda = g^2 N_c$ in $\symff$ does not run and is independent of the temperature.  In the weak-coupling limit, the thermodynamics of $\symff$ has been calculated through order $\lambda^2$ with the result being~\cite{Fotopoulos:1998es,Kim:1999sg,VazquezMozo:1999ic,Nieto:1999kc,Du:2021jai,Andersen:2021bgw} 
\bqa\label{weaexp}
&& \frac{\mathcal{F}}{\mathcal{F}_{\textrm{ideal}}}= \frac{\mathcal{S}}{\mathcal{S}_{\textrm{ideal}}} = 1-\frac{3}{2} \frac{\lambda}{\pi^2} +  \left( 3+\sqrt{2} \right) \left(\frac{\lambda}{\pi^2}\right)^{3/2}   \nonumber \\ && \hspace{0.6cm} + \bigg[ -\frac{21}{8} -\frac{9\sqrt{2}}{8} + \frac{3}{2} \gamma_E + \frac{3}{2}\frac{\zeta'(-1)}{\zeta(-1)} \nonumber \\ && \hspace{2.5cm} -\frac{25}{8} \log 2 + \frac{3}{2}\log \frac{\lambda}{\pi^2} \bigg] \left(\frac{\lambda}{\pi^2}\right)^2 ,
\eqa
where $\mathcal{F}_{\textrm{ideal}} = - d_A \pi^2T^4/6$ is the ideal or Stefan-Boltzmann limit of the free energy and $\mathcal{S}_{\textrm{ideal}} = 2 d_A \pi^2T^3/3$, with $d_A = N_c^2 -1$ being the dimension of the adjoint representation and $\zeta(z)$ being the Riemann zeta function.  The ratios of the free energy and entropy density to their corresponding ideal limit are the same owing to the fact that the `t Hooft coupling is temperature independent.  Note, importantly that the weak-coupling expression \eqref{weaexp} is valid for all $N_c$.  

In the strong-coupling limit, the behavior of the $\symff$ free energy has been computed using the AdS/CFT correspondence. Most information is known about the large-$N_c$ limit where one has~\cite{Gubser:1998nz}
\beq\label{stro}
\frac{\mathcal{F}}{\mathcal{F}_{\textrm{ideal}}}= \frac{\mathcal{S}}{\mathcal{S}_{\textrm{ideal}}} =\frac{3}{4}\bigg[1+\frac{15}{8}\zeta(3)\lambda^{-3/2} + \mathcal{O}(\lambda^{-2}) \bigg] .
\eeq

One issue that must be faced when using strict weak- or strong-coupling expansions is that they might have poor convergence as additional orders are included in the expansion or may not converge at all.
This is a known issue with the weak-coupling expansion of QCD thermodynamics, in which case the series seems to converge only for $T \gtrsim 10^5$ GeV.  This poor convergence of strict perturbation theory motivated research into methods for reorganizing the weak-coupling expansion in order to improve its convergence as one goes to higher loop order.  In the context of QCD, the methods used have included the $\Phi$-derivable method \cite{Luttinger:1960ua,Baym:1962sx,Cornwall:1974vz,Freedman:1976ub} and the hard-thermal-loop perturbation theory (HTLpt) reorganization~\cite{Andersen:1999fw,Andersen:1999sf,Andersen:1999va}.  

Despite its appeal, a fundamental issue with the $\Phi$-derivable method is that it is not manifestly gauge invariant, with gauge parameter dependence appearing at the same order in $\lambda$ as the series truncation when evaluated off the stationary point and at twice the order in $\lambda$ when evaluated at the stationary point \cite{Arrizabalaga:2002hn,Blaizot:1999ip,Andersen:2004re}.  The HTLpt approach, on the contrary, is manifestly gauge invariant due to the fact that the HTL effective action used as the starting point is gauge invariant by construction.  HTLpt has been used to improve the convergence of weak coupling calculations of the free energy in scalar field theories~\cite{Andersen:2000yj,Andersen:2001ez,Andersen:2008bz}, QED~\cite{Andersen:2009tw}, and QCD up to three-loop order at finite temperature and chemical potential \cite{Andersen:2009tc,Andersen:2010ct,Andersen:2010wu,Andersen:2011sf,Andersen:2011ug,Haque:2013sja,Haque:2014rua}.  Based on its success in QCD applications, in Ref.~\cite{Du:2020odw} we applied this method to $\symff$, however, in this prior work canonical dimensional regularization (DRG) \cite{Ashmore:1972uj,Bollini:1972ui,THOOFT1972189} was used to regulate divergences generated during the calculation in the same manner as was done in QCD.  

One issue with this prior work is that the use of canonical DRG breaks supersymmetry because in DRG the size of the bosonic representation depends on the dimensional regulation parameter $\epsilon$.  In this paper we address this issue by using a regularization scheme called regularization by dimensional reduction (RDR).  The RDR method was introduced by Siegel \cite{Siegel:1979wq} and is a modified version of dimensional regularization \cite{Brink:1976bc,Gliozzi:1976qd} which manifestly preserves gauge invariance, unitarity, and supersymmetry \cite{Avdeev:1982xy,CAPPER1980479}.  In Refs.~\cite{Du:2021jai,Andersen:2021bgw} this method was used to compute the $\lambda^2$ and $\lambda^2 \log\lambda$ coefficients in Eq.~\eqref{weaexp}.  Herein we will compute the order $\lambda$, $\lambda^{3/2}$, and $\lambda^2 \log\lambda$ coefficients using two-loop HTLpt and demonstrate that these coefficients are scheme independent.  We will additionally demonstrate that the order $\lambda^2$ coefficient is regularization-scheme dependent; however, this is somewhat expected, since this coefficient is beyond the strict perturbative accuracy of a two-loop calculation.  Importantly, we find that the coefficient of $\lambda^2\log\lambda$ is exactly the same as obtained in the prior two-loop DRG HTLpt calculation \cite{Du:2020odw}, resummation using the Arnold-Zhai method \cite{Du:2021jai}, and resummation using effective field theory methods \cite{Andersen:2021bgw}.  This firmly establishes the existence of logarithms in the weak-coupling expansion of $\symff$ thermodynamics and provides confidence in the computed coefficient.

Note that since the high-order terms in the expansion of the free energy are unknown, it not possible to determine if the finite temperature perturbative series for the free energy has a finite radius of convergence.  We note, however, that even if the perturbative series has zero radius of convergence (an asymptotic series), it is possible to apply variational perturbation theory methods such as HTLpt to improve the convergence of successive loop approximations.  In cases where all orders expansions of quantities are known, e.g., the ground state energy of a zero temperature anharmonic oscillator \cite{Bender:1969si,Bender:1973rz}, it has been shown that by using variational perturbation theory one can even self-consistently obtain the strong coupling limit coefficients from a divergent weak-coupling expansion \cite{Janke:1995zz,Kleinert:1995hc}.  

In the case of $\symff$, our recent calculations through order $\lambda^2$ \cite{Du:2021jai,Andersen:2021bgw} suggest that the perturbative expansion of $\symff$ thermodynamics has a finite and potentially large radius of convergence.  The resummed results obtained in this paper show that HTLpt can be used to improve the convergence of successive approximations to the $\symff$ free energy. As will we demonstrate, the two-loop HTLpt resummed result, although having strict perturbative accuracy of order $\lambda^{3/2}$, reproduces the order $\lambda^2$ result to within 2\% for $\lambda \lesssim 2$  (see Fig.~\ref{HTLNLA1RDR} below).

Finally, we emphasize that the scheme dependence discussed herein is related to whether or not one uses a supersymmetry-preserving regularization scheme.  This is a fundamental symmetry requirement and, hence, the RDR scheme is better suited to this problem.  In the context of QED and QCD there exists a more general scheme dependence which stems from the choice of the method of regularization.  The standard scheme for renormalizing QCD is the $\overline{\text{MS}}$ (modified minimal subtraction) scheme \cite{Bardeen:1978yd}, which is related to the $\text{MS}$ (minimal subtraction) scheme \cite{tHooft:1973mfk,Weinberg:1973xwm} through a rescaling of the $\text{MS}$ scale by $(e^{\gamma_E}/4\pi)^\epsilon$ where $\gamma_E$ is the Euler-Mascheroni constant.  As a result of this relationship, one can connect quantities computed in the two schemes in a straightforward manner using the running coupling itself~\cite{Bardeen:1978yd}.  We note importantly, however, that physical observables such as scattering rates and the free energy are scheme independent~\cite{Bardeen:1978yd,Stevenson:1982wn,Lepage:1989hf}.  In strict perturbation theory, renormalization scheme invariance of observables can be established order-by-order in the coupling, however, the parameters in the theory which are not directly observable, such as the running coupling constant, are in general scheme dependent.  When computing observables the scheme-dependence of the parameters is compensated for by the scheme-dependence of the coefficients in the perturbative expansion thereby ensuring renormalization group invariance~\cite{Bardeen:1978yd,Stevenson:1982wn,Lepage:1989hf}.

In $\symff$ an analogous scheme dependence does not occur because the theory is conformal and the coupling does not run.  As a result, the perturbative coefficients are renormalization group invariants and hence, scheme independent in the QED/QCD sense. This is evidenced by Eq.~\eqref{weaexp} which is scale invariant through the perturbative accuracy determined. Finally, we note that HTLpt goes beyond strict perturbation theory through an all orders resummation in the soft sector.  As a result, a residual scale dependence can remain in HTLpt when truncating at finite loop order; however, as one extends the HTLpt calculation to higher loop order, the scale dependence is systematically pushed to higher orders in the `t Hooft coupling, resulting in mathematically unique predictions for the fully determined perturbative coefficients.

The structure of our paper is as follows.  We begin with a brief introduction to the basics of $\symff$ in Sec.~\ref{chap01sec02}. In section \ref{subsec1}, we present a summary of HTLpt applied to $\symff$.  We list the terms which change in the high temperature expansion when going from DRG to RDR in Sec.~\ref{chap0401}. Based on these results, we present the complete expressions for the leading- (LO) and next-to-leading order (NLO) thermodynamic potentials in the RDR scheme in Sec.~\ref{chap0402}. In Sec.~\ref{chap0403}, we present our numerical results for the RDR HTLpt-resummed NLO $\symff$ scaled thermodynamic functions and compare to our previous results obtained using the DRG scheme.  We also compare to strict perturbative expression for the scaled thermodynamic functions through order $\lambda^2$ which were obtained using the RDR scheme and to a generalized Pad\'{e} approximant based on this result and the corresponding result in the large-$N_c$ strong coupling limit.  In Sec.~\ref{conclusions} we present our conclusions and an outlook for the future.

{\em Notation:} We use lower-case letters for Minkowski space four-vectors, e.g., $p$, and upper-case letters for Euclidean space four-vectors, e.g., $P$.  We use the mostly minus convention for the metric.


\section{The basics of the $\symff$ theory}
\label{chap01sec02}

In $\symff$ all fields belong to the adjoint representation of the $SU(N_c)$ gauge group. The definition of gauge field is the same as QCD and $A_\mu$ can be expanded as \mbox{$A_\mu=A_\mu^a t^a$}, with real coefficients $A_\mu^a$, and Hermitian color generators $t^a$ in the adjoint representation which satisfy
\bqa\label{A}
 [t^a,t^b]=i f_{abc}t^c   \quad \textrm{and} \quad \textrm{Tr}(t^a t^b)=\frac{1}{2}\delta^{ab} \, ,
\eqa
where $a,b=1, \cdots ,N_c^2-1$ and the group structure constants $f_{abc}$ are real and completely antisymmetric. 

For the fermionic fields, the massless two-component Weyl fermions $\psi$ in four dimensions can be converted into four-component Majorana fermions \cite{Quevedo:2010ui,bertolini2015lectures,Yamada:2006rx,DHoker:1999yni,Kovacs:1999fx}
\bqa
\psi \equiv  \begin{pmatrix} \psi_\alpha\\  \bar{\psi}^{\dot{\alpha}} \end{pmatrix}  \quad\quad \textrm{and} \quad\quad \bar{\psi} \equiv  \begin{pmatrix} \psi^\alpha & \bar{\psi}_{\dot{\alpha}} \end{pmatrix} ,
\eqa
where $\alpha=1,2$ and the Weyl spinors satisfy $\bar{\psi}^{\dot{\alpha}}\equiv [\psi^\alpha]^\dagger$. The conjugate Majorana spinor $\bar{\psi}$ is not independent, but is related to $\psi$ via the Majorana condition $\psi=C\bar{\psi}$, where $C=\begin{pmatrix}\begin{smallmatrix} \epsilon_{\alpha\beta} & 0 \\ 0 & \epsilon^{\dot{\alpha}\dot{\beta}}\end{smallmatrix} \end{pmatrix}$ is the charge conjugation operator with $\epsilon_{02}=-\epsilon_{11}\equiv-1$. We will use Latin indices $i,j=1,2,3,4$ to label the four Majorana fermions, with $\psi_i$ denoting each bispinor.  Since the fermions are in the adjoint representation, one can expand $\psi_i=\psi_i^a t^a$, where the coefficients $\psi_i^a$ are four-component Grassmann-valued Majorana spinors.

In addition to the gauge field and Majorana spinors, there are six independent real scalar fields which are represented by a multiplet
\bqa\label{la}
 \Phi\equiv (X_1,Y_1,X_2,Y_2,X_3,Y_3) \, ,
\eqa
where $X_{\texttt{p}}$ and $Y_{\texttt{q}}$ are Hermitian, with ${\texttt{p,q}}=1,2,3$. $X_{\texttt{p}}$ and $Y_{\texttt{q}}$ denote scalar and pseudoscalar fields, respectively. We will use a capital Latin index $A$ to denote components of the vector $\Phi$. Therefore $\Phi_A$, $X_{\texttt{p}}$, and $Y_{\texttt{q}}$ can be expanded as $\Phi_A=\Phi_A^a t^a$, with $A=1, \cdots ,6$ or alternatively, $X_{\texttt{p}}=X_{\texttt{p}}^at^a$ and $Y_{\texttt{q}}=Y_{\texttt{q}}^at^a$.

The Minkowski-space Lagrangian density for $\symff$ can be expressed as
\bqa\label{lag}
&& \mathcal{L}_{\symff} = \textrm{Tr}\bigg[{-}\frac{1}{2}G_{\mu\nu}^2+(D_\mu\Phi_A)^2+i\bar{\psi}_i {\displaystyle{\not} D}\psi_i \nonumber \\ && \hspace{6mm} -\frac{1}{2}g^2(i[\Phi_A,\Phi_B])^2 - i g \bar{\psi}_i\big[\alpha_{ij}^{\texttt{p}} X_{\texttt{p}}+i \beta_{ij}^{\texttt{q}}\gamma_5Y_{\texttt{q}},\psi_j\big] \bigg] \nonumber \\ && \hspace{6mm} +\mathcal{L}_{\textrm{gf}}+\mathcal{L}_{\textrm{gh}}+\Delta\mathcal{L}_{\textrm{SYM}} \, ,
\eqa
where $\mu, \nu=0,1,2,3$ and $\alpha^{\texttt{p}}$ and $\beta^{\texttt{q}}$ are $4\times 4$ matrices that satisfy
\beq\label{lag1}
\{\alpha^{\texttt{p}},\alpha^{\texttt{q}}\}=-2\delta^{\texttt{pq}} \, , \;\;\;  \{\beta^{\texttt{p}},\beta^{\texttt{q}} \}=-2\delta^{\texttt{pq}} \, , \;\;\;  [\alpha^{\texttt{p}},\beta^{\texttt{q}}]=0 \, .
\eeq
The matrices $\alpha$ and $\beta$ satisfy $\alpha_{ik}^{\texttt{p}}\alpha_{kj}^{\texttt{p}}=-3\delta_{ij}$ and $\beta_{ij}^{\texttt{q}}\beta_{ji}^{\texttt{p}}=-4\delta^{\texttt{pq}}$, with $\delta_{ii}=4$ for the four Majorana fermions and $\delta^{\texttt{pp}}=3$ for each set of three scalars.

To quantize the theory, gauge-fixing and ghost terms should be added to the Lagrangian density. In general covariant gauge, their forms are the same as in QCD,
\bqa\label{gf}
\mathcal{L}_{\textrm{gf}}^{\symff} &=& -\frac{1}{\xi}\textrm{Tr}\big[(\partial^\mu A_\mu)^2 \big],  \nonumber \\
\mathcal{L}_{\textrm{gh}}^{\symff} &=&-2\textrm{Tr}\big[\bar{\eta}\,\partial^\mu \! D_\mu \eta \big] ,
\eqa
with $\xi$ being the gauge parameter.

\section{ HTL\lowercase{pt} for $\symff$ theory}
\label{subsec1}

HTLpt provides a way to incorporate plasma effects into resummed perturbative calculations while maintaining explicit gauge invariance. In HTLpt the underlying theory is reorganized by adding and subtracting the HTL effective action to the vacuum action.  The addition/subtraction of the HTLpt action generates effective propagators and vertices which are functions of energy and momentum. The HTLpt method has been applied to QCD to three-loop order in Refs.~\cite{Andersen:1999fw,Andersen:1999sf,Andersen:1999va,Andersen:2002ey,Andersen:2003zk,Andersen:2009tc,Andersen:2010ct,Andersen:2010wu,Andersen:2011sf,Andersen:2011ug,Haque:2012my,Haque:2013qta,Haque:2013sja,Haque:2014rua,Andersen:2015eoa,Haque:2020eyj}.

The HTLpt reorganization of $\symff$ can be obtained in the same manner as in QCD by introducing an expansion parameter $\delta$, treating it as a formal expansion parameter, expanding around $\delta=0$ to a fixed order, and then setting $\delta=1$ in the end.  The HTL reorganized Lagrangian density for $\symff$ can be written as
\beq\label{dla}
\mathcal{L}^{\textrm{shifted}}_{\symff} = (\mathcal{L}_{\symff}+\mathcal{L}^{\textrm{HTL}}_{\symff})|_{g\rightarrow\sqrt{\delta}g}+\Delta\mathcal{L}^{\textrm{HTL}}_{\symff} \, .
\eeq
The HTL improvement term $\mathcal{L}^{\textrm{HTL}}_{\symff}$ is
\bqa\label{htl}
\mathcal{L}^{\textrm{HTL}}_{\symff}&=& -\frac{1}{2}(1-\delta)m_D^2 \, \textrm{Tr}\bigg (G_{\mu \alpha} \bigg \langle\frac{y^\alpha y^\beta}{(y\cdot D)^2}\bigg \rangle_{\hat{\textbf{y}}} G_\beta^\mu \bigg )\nonumber
\\&&\hspace{0.6cm}
+ (1-\delta)i m_q^2 \, \textrm{Tr}\bigg (\bar{\psi}_j \gamma^\mu \bigg \langle \frac{y^\mu}{y\cdot D}   \bigg \rangle_{\hat{\textbf{y}}} \psi_j  \bigg ) \nonumber
\\&&\hspace{1.5cm}-(1-\delta) M_D^2 \, \textrm{Tr} (\Phi_A^2 ) \, ,
\eqa
where $y^\mu=(1,\hat{\textbf{y}})$ is a light-like four vector, $\langle\cdots \rangle_{\hat{\textbf{y}}}$ represents an average over the direction of $\hat{\textbf{y}}$ defined in App.~\ref{fmr}, the index $j \in \{1  \ldots 4\}$ labels the four Majorana fermions, and the index $A \in \{ 1 \ldots 6\}$ labels the six independent real-valued scalars. The parameters $m_D$ and $M_D$ are the electric screening masses for the gauge field and the adjoint scalar fields, respectively. The parameter $m_q$ is the induced finite temperature quark mass. We note that, in general, the gluon mass in RDR is not equal to $-g_{\mu\nu}\Pi^{\mu\nu}$, where $\Pi^{\mu\nu}$ is the gluonic self energy and care must be taken due to this.  This stems from the necessity of employing integration by parts during the calculation of the gluon self energy. 

In the RDR scheme, all momentum-space integrals will be evaluated in $d=4-2\epsilon$ dimensions while the size of the gauge field and fermionic field representations will be taken to be integer valued, and correspond to $D=4$ dimensional fields. Since, as we will demonstrate, all ultraviolet divergences generated in the calculation of HTL-thermodynamics are canceled by the HTLpt counterterms independently of the value of the gluon, quark, and scalar mass parameters, we need to only consider the leading-order contributions in the momentum-space regulation parameter $\epsilon$.

The HTLpt reorganization of QCD and $\symff$ generates new ultraviolet (UV) divergences compared to the vacuum Lagrangian.  In QCD, these divergences can be eliminated using the counterterm Lagrangian $\Delta\mathcal{L}^{\textrm{HTL}}_{\textrm{QCD}}$ and the thermodynamic potential at the two-loop level can be renormalized by using a simple counterterm Lagrangian $\Delta\mathcal{L}^{\textrm{HTL}}_{\textrm{QCD}}$ which includes vacuum energy and mass counterterms~\cite{Andersen:2002ey,Andersen:2003zk}. Although not proven at arbitrary loop order, it has been explicitly demonstrated that one can renormalize the HTLpt thermodynamic potential through three-loop order using only vacuum, gluon thermal mass, quark thermal mass, and gauge coupling constant counterterms~\cite{Andersen:2011sf,Haque:2014rua}. The same method can be used in $\symff$.

We find that in $\symff$ the vacuum counterterm $\Delta_0 \mathcal{E}_0$, which is the leading order counterterm in the $\delta$ expansion of the vacuum energy $\mathcal{E}_0$, can be obtained by calculating the free energy to leading order in $\delta$, and the next-to-leading-order contribution $\Delta_1 \mathcal{E}_0$ can be obtained by expanding $\Delta \mathcal{E}_0$ to linear order in $\delta$. As a result, we find that in the RDR scheme the counterterm $\Delta \mathcal{E}_0$ has the form
\bqa\label{dc0}
\Delta \mathcal{E}_0 & =& \bigg(\frac{d_A}{128\pi^2 \epsilon} +O(\lambda\delta ) \bigg)(1-\delta)^2m_D^4 \nonumber
\\& +& \bigg(\frac{3d_A}{32\pi^2 \epsilon}+O(\lambda\delta )\bigg)(1-\delta)^2 M_D^4 \, .
\eqa
Comparing to the DRG HTLpt result obtained in Ref.~\cite{Du:2020odw}, one sees that Eq.~(\ref{dc0}) is precisely the same as obtained therein.  

To calculate the NLO HTLpt-improved free energy one needs to expand the partition function to order $\delta$.  Correspondingly, to cancel the remaining UV divergences we need the counterterms $\Delta\mathcal{E}_0$, $\Delta m_D^2$, $\Delta m_q^2$, and $\Delta M_D^2$ to order $\delta$.  In order to remove all UV divergences which appear at two-loop level, the mass counterterms required are
\bqa\label{dmass}
\Delta m_D^2&=&\bigg(\frac{1}{16\pi^2\epsilon}\lambda\delta +O( \lambda^2\delta^2) \bigg)(1-\delta) m_D^2  \, ,  \nonumber \\
\Delta M_D^2&=&\bigg(\frac{3}{8\pi^2 \epsilon}\lambda\delta + O( \lambda^2\delta^2) \bigg)(1-\delta)  M_D^2 \, ,   \nonumber \\
\Delta m_q^2&=&\bigg(-\frac{1}{\pi^2 \epsilon}\lambda\delta + O( \lambda^2\delta^2) \bigg)(1-\delta) m_q^2 \, .
\eqa
Once again we find that the mass counterterms are precisely the same as the ones necessary to remove the HTLpt divergences in DRG HTLpt~\cite{Du:2020odw}.

To calculate HTLpt-improved physical observables in $\symff$ we use the same method as in QCD, namely expanding the path-integral in powers of $\delta$, truncating at some specified order, and then setting $\delta=1$. The results for physical observables will depend on $m_D$, $M_D$, and $m_q$ which are functions of $T$ and $\lambda$.  These parameters are fixed in HTLpt by minimizing the free energy. If we use $\Omega^{\textrm{RDR}}_N(T,\lambda, m_D, M_D, m_q,  \delta)$ to represent the thermodynamic potential expanded to $N$-th order in $\delta$, then the corresponding variational prescription is
\bqa\label{gap}
\frac{\partial}{\partial m_D}\Omega^{\textrm{RDR}}_N(T, \lambda, m_D, M_D, m_q, \delta=1)&=&0 \, ,    \nonumber \\
\frac{\partial}{\partial M_D}\Omega^{\textrm{RDR}}_N(T, \lambda, m_D, M_D, m_q, \delta=1)&=&0 \, ,     \nonumber \\
\frac{\partial}{\partial m_q}\Omega^{\textrm{RDR}}_N(T, \lambda, m_D, M_D, m_q, \delta=1)&=&0 \, .
\eqa
These three equations are called the \textit{gap equations}. The free energy can be obtained by evaluating the thermodynamic potential at the solution to the gap equations, and other thermodynamic functions can be obtained from the free energy and its derivatives with respect to $T$.

In the following section, we calculate the thermodynamic potential through order $\lambda^{5/2}$, which can be expressed as an expansion in powers of $m_D/T$, $m_q/T$, and $M_D/T$. Instead of repeating all results from our DRG calculation, we list all terms which change when going from DRG to RDR. As mentioned above, at order $\delta$, all divergences  in the two-loop thermodynamic potential can be removed by the HTLpt vacuum and mass counterterms which are scheme independent. 

\section{High-temperature expansion using the RDR scheme}
\label{chap0401}

In this section, we list only the contributions which change when going from the DRG to RDR scheme. Since the Feynman rules and diagrams needed are the same as in Ref.~\cite{Du:2020odw}, the high-temperature forms are the same as in this reference up to the dimension of the gluon tensor and metric tensor.  The labels on each Feynman diagram contribution below correspond to the diagrams shown in Ref.~\cite{Du:2020odw}.

\subsection{One-loop sum-integrals in the RDR scheme}

Based on the above knowledge, at one-loop order, only the hard contribution from the gluonic one-loop free energy and two-loop HTL counterterm are modified in the RDR scheme. The remaining contributions come from fermions and scalars and remain unchanged compared to the DRG HTLpt result.

\subsubsection{RDR modified gluon hard contributions }

The form of the hard contribution to the one-loop gluon free energy expanded to second order in $m_D^2$ has been given in \cite{Andersen:2002ey}, and can be written in the RDR scheme as
\bqa\label{ghRDR}
&& \mathcal{F}_{g}^{(h), \textrm{RDR}} = \frac{D-2}{2}b_0 + \frac{1}{2}m_D^2 b_1 - \frac{1}{4(D-2)}m_D^4 b_2 \nonumber \\ && \hspace{0.5cm} - \frac{1}{4(D-2)}m_D^4 \sumint_{P} \bigg[  - 2\frac{1}{p^2 P^2}  -2 (D-1)\frac{1}{p^4} {\mathcal T}_P \nonumber \\ && \hspace{2.5cm}+ 2\frac{1}{p^2 P^2}{\mathcal T}_P +(D-1) \frac{1}{p^4} ({\mathcal T}_P)^2 \bigg],
\eqa
where ${\mathcal T}_P$ is given in Eq.~(\ref{t00x1}), which enters into the HTL gluon self-energy, and
\bqa\label{formb0bn}
b_0 & \equiv & \sumint_P \log P^2 = -\frac{\pi^2}{45} T^4\, ,  \label{forlumab0} \\
b_n & \equiv & \sumint_{P} \frac{1}{P^{2n}} \, ,   \quad n \geq 1\, .\label{forluma1}
\eqa
Setting $D=4$, using Eq.~(\ref{formb0bn}), and the formulas in App.~\ref{IntegralHTLpt}, \eqref{ghRDR} reduces to
\bqa\label{ghRDR1}
&& \mathcal{F}_{g}^{(h), \textrm{RDR}} = -\frac{\pi^2}{45}T^4 +\frac{1}{24}\bigg[ 1+ \bigg(2+2\frac{\zeta^{'}(-1)}{\zeta(-1)}   \bigg)\epsilon   \bigg] \nonumber \\ && \hspace{0.5cm} \times
 \bigg( \frac{\mu}{4\pi T}   \bigg)^{2\epsilon} m_D^2 T^2  - \frac{1}{128\pi^2}\bigg(\frac{1}{\epsilon} + 2\gamma+\frac{2\pi^2}{3} \nonumber \\ && \hspace{2.3cm}-\frac{16}{3}  -\frac{8 \log 2}{3} \bigg)\bigg( \frac{\mu}{4\pi T}   \bigg)^{2\epsilon}m_D^4\,.
\eqa

The form of the hard contribution to the two-loop gluon HTL counterterm has been given in \cite{Andersen:2002ey}, and can be expressed as
\bqa\label{gcthRDR}
&& \mathcal{F}_{gct}^{(h), \textrm{RDR}} = -\frac{1}{2} m_D^2 b_1 + \frac{1}{2(D-2)}m_D^4 b_2 \nonumber \\ && \hspace{0.8cm} + \frac{1}{2(D-2)}m_D^4 \, \sumint_{P} \bigg[ - 2\frac{1}{p^2 P^2}  -2 (D-1)\frac{1}{p^4} {\mathcal T}_P \nonumber \\ &&\hspace{1.8cm} + 2\frac{1}{p^2 P^2}{\mathcal T}_P +(D-1) \frac{1}{p^4} ({\mathcal T}_P)^2 \bigg] .
\eqa
Setting $D=4$, using Eq.~(\ref{formb0bn}), and the formulas in App.~\ref{IntegralHTLpt}, \eqref{gcthRDR} reduces to
\bqa\label{gcthRDR1}
&& \mathcal{F}_{gct}^{(h), \textrm{RDR}}  =  -\frac{1}{24}m_D^2 T^2+\frac{1}{64\pi^2}\bigg[ \frac{1}{\epsilon}+2\gamma+\frac{2\pi^2}{3} \nonumber \\ && \hspace{2.7cm} -\frac{16}{3}   - \frac{8\log 2}{3}   \bigg] \bigg( \frac{\mu}{4\pi T}   \bigg)^{2\epsilon}m_D^4\,.
\eqa
We find, as can be expected on general grounds, that the only difference between the result of the one-loop gluonic contributions in the RDR scheme and the DRG scheme is the finite contribution which is proportional to $m_D^4$.

\subsection{Two-loop sum-integrals in the RDR scheme}

Considering the two-loop contributions in the RDR scheme, the soft-soft $(ss)$ contribution from all diagrams and the $\mathcal{F}_{3qs}^{(hh)}$ contribution are unchanged. All other contributions are modified due to the dimension of the gluon tensor and metric tensor changing from $4-2\epsilon$ to 4.

\subsubsection{Contributions from the $(hh)$ region}

The form of the $(hh)$ contributions to the two-loop gluonic free energy expanded to first order in $m_D^2$ has been given in \cite{Andersen:2002ey}, and can be written in the RDR scheme as
\begin{widetext}
\bqa\label{ghhRDR}
&& \mathcal{F}_{3g+4g+gh}^{(hh),\textrm{RDR}} = \frac{1}{4}(D-2)^2 b_1^2 - (D-2) \frac{1}{2}m_D^2 b_1 b_2  + \frac{1}{4}m_D^2 \sumint_{PQ} \bigg\{2 (D-3)\frac{1}{P^2Q^2q^2} +2 \frac{1}{P^2Q^2R^2}  \nonumber \\ &&  + (D+1)\frac{1}{P^2Q^2r^2}  -2(D-1)\frac{P\cdot Q}{P^2Q^2 r^4} - 4(D-1)  \frac{q^2}{P^2Q^2 r^4}  + 4 \frac{q^2}{P^2Q^2R^2r^2}  -2 (D-2)\frac{1}{P^2Q^2 q^2}{\mathcal T}_Q \nonumber \\ && -D \frac{1}{P^2Q^2 r^2}{\mathcal T}_R  + 4 (D-1) \frac{q^2}{P^2Q^2r^4} {\mathcal T}_R   + 2(D-1)\frac{P\cdot Q}{P^2Q^2 r^4} {\mathcal T}_R  \bigg\}   \, .
\eqa
\end{widetext}
Setting $D=4$, using Eq.~(\ref{formb0bn}), and the formulas in App.~\ref{IntegralHTLpt}, \eqref{ghhRDR} reduces to
\bqa\label{ghhRDR1}
\mathcal{F}_{3g+4g+gh}^{(hh),\textrm{RDR}} &=& \frac{1}{144} T^4-\frac{29}{4608\pi^2}\bigg( \frac{1}{\epsilon}+ 5.751206124  \bigg)\nonumber \\ &\times& \bigg( \frac{\mu}{4\pi T}   \bigg)^{4\epsilon} m_D^2T^2\,.
\eqa

The form of the $(hh)$ contribution to $\mathcal{F}_{3qg}$ and $\mathcal{F}_{4qg}$ expanded to first order in $m_D^2$ can be obtained from Ref.~\cite{Andersen:2003zk}, and in the RDR scheme become
\begin{widetext}
\bqa\label{qhhRDR}
&& \mathcal{F}_{3qg+4qg}^{(hh),\textrm{RDR}}=  2(D-2)[f_1^2-2b_1 f_1] + 4m_D^2 \bigg\{ b_2 f_1  + \sumint_{P\{Q\}} \bigg[ \frac{1}{p^2P^2Q^2} {\mathcal T}_P  -\frac{D-3}{D-2}\frac{1}{p^2P^2Q^2}\bigg] \bigg\}  + 2m_D^2 \sumint_{\{PQ\}}   \bigg[ \frac{D}{D-2} \nonumber \\ && \hspace{0.2cm} \times \frac{1}{P^2Q^2r^2}  - \frac{4(D-1)}{D-2}\frac{q^2}{P^2Q^2r^4} - \frac{2(D-1)}{D-2} \frac{P\cdot Q}{P^2Q^2r^4}\bigg]{\mathcal T}_R  + 2m_D^2 \sumint_{\{PQ\}} \bigg[ \frac{4-D}{D-2} \frac{1}{P^2Q^2R^2}  +\frac{2(D-1)}{D-2}\frac{P\cdot Q}{P^2Q^2r^4} \nonumber \\ && \hspace{0.2cm} - \frac{D+1}{D-2}\frac{1}{P^2Q^2r^2} + \frac{4(D-1)}{D-2}  \frac{q^2}{P^2Q^2r^4}  - \frac{4}{D-2}\frac{q^2}{P^2Q^2R^2r^2}\bigg] + 4 m_q^2 (D-2)\sumint_{\{PQ\}} \bigg[ \frac{1}{P^2Q^2Q_0^2}  + \frac{p^2-r^2}{P^2q^2Q_0^2R^2} \bigg]{\mathcal T}_Q  \nonumber \\ && \hspace{-0.3cm} + 4 m_q^2 (D-2) \bigg[ 2b_1 f_2  - \sumint_{P\{Q\}} \frac{1}{P^2Q^2Q_0^2} {\mathcal T}_Q \bigg]  + 4 m_q^2 (D-2) \bigg\{ -2 f_1f_2  + \sumint_{\{PQ\}} \bigg[ \frac{D-6}{D-2}\frac{1}{P^2Q^2R^2}  + \frac{r^2-p^2}{P^2Q^2R^2 q^2}\bigg]  \bigg\}   \,,
\eqa
\end{widetext}
where $f_n$ is defined as
\bqa\label{forlumaf1}
f_n \equiv \sumint_{\{P\}} \frac{1}{P^{2n}} =(2^{2n+1-d}-1)b_n \, ,  \quad n \geq 1 \,.
\eqa
Setting $D=4$, using Eqs.~(\ref{formb0bn}) and (\ref{forlumaf1}), and formulas in App.~\ref{IntegralHTLpt}, \eqref{qhhRDR} reduces to
\bqa\label{qhhRDR1}
&& \mathcal{F}_{3qg+4qg}^{(hh),\textrm{RDR}} = \frac{5}{144}T^4-\frac{7}{1152\pi^2}\bigg[ \frac{1}{\epsilon}+ 0.1415352337  \bigg] \nonumber \\ && \hspace{0.5cm}\times \bigg( \frac{\mu}{4\pi T}   \bigg)^{4\epsilon} m_D^2T^2 + \frac{1}{16\pi^2}\bigg[ \frac{1}{\epsilon}+ 9.96751112  \bigg] \nonumber \\ && \hspace{1.6cm}\times \bigg( \frac{\mu}{4\pi T}   \bigg)^{4\epsilon}m_q^2T^2 \,.
\eqa

The form of the $(hh)$ contributions to $\mathcal{F}_{4s}$, $\mathcal{F}_{3gs}$ and $\mathcal{F}_{4gs}$ are
\bqa\label{f4shhRDR}
&& \mathcal{F}_{4s+3gs+4gs}^{(hh),\textrm{RDR}}= 3(D+1)b_1^2 -3(4+D) M_D^2 b_1 b_2  + M_D^2 \nonumber \\ && \times\sumint_{PQ}\frac{6}{P^2Q^2R^2}  -3 m_D^2 b_1 b_2  + m_D^2 \frac{3(D-3)}{D-2}b_1  \sumint_{P}\frac{1}{p^2P^2}\nonumber \\ && 
+\frac{m_D^2}{D-2}\sumint_{PQ}\bigg\{\frac{3(1+D)}{2p^2Q^2R^2}-\frac{3}{P^2Q^2R^2}
 -\frac{6(D-1)q^2}{p^4Q^2R^2} \nonumber \\ && +\frac{6q^2}{p^2P^2Q^2R^2}+\frac{3(D-1)(Q\cdot R)}{p^4Q^2R^2}+\bigg[\frac{3(2-D)}{p^2P^2Q^2} \nonumber \\
 && \hspace{-0.1cm} -\frac{3D}{2p^2Q^2R^2}+\frac{6(D-1)q^2}{p^4Q^2R^2}-\frac{3(D-1)(Q\cdot R)}{p^4Q^2R^2}    \bigg] {\mathcal T}_P \bigg\}  \, .
\eqa
Setting $D=4$, using Eq.~(\ref{formb0bn}), and the formulas in App.~\ref{IntegralHTLpt}, \eqref{f4shhRDR} reduces to
\bqa\label{f4shhRDR1}
&& \mathcal{F}_{4s+3gs+4gs}^{(hh),\textrm{RDR}}= \frac{5}{48}T^4-\frac{1}{8\pi^2}\bigg( \frac{1}{\epsilon}+2\gamma+ 5.97010745  \bigg) \nonumber \\ && \hspace{0.3cm}\times \bigg( \frac{\mu}{4\pi T}   \bigg)^{4\epsilon}M_D^2T^2-\frac{29}{1536\pi^2} \bigg( \frac{1}{\epsilon}+ 5.751206124  \bigg) \nonumber \\ && \hspace{1.3cm}\times \bigg( \frac{\mu}{4\pi T}   \bigg)^{4\epsilon}m_D^2T^2\,.
\eqa
%

\subsubsection{Contributions from the $(hs)$ region}

The form of the $(hs)$ contribution to the two-loop gluon free energy expanded to first order in $m_D^2$ was given in \cite{Andersen:2002ey} and in the RDR scheme can be written as
\bqa\label{ghsRDR}
&& \mathcal{F}_{3g+4g+gh}^{(hs),\textrm{RDR}} =  \frac{T}{2} \int_{\textbf{p}} \frac{1}{p^2+m_D^2}  \bigg\{  -(D-2)b_1 + 2(D-2) \nonumber \\ && \hspace{0.5cm} \times \sumint_{Q} \frac{q^2}{Q^4}  \bigg\}  +  m_D^2 T  \int_{\textbf{p}} \frac{1}{p^2+m_D^2} \bigg\{    -(D-5)b_2 \nonumber \\ && \hspace{0.8cm} +\sumint_{Q} \bigg[ \frac{(D-2)(D+1)}{D-1} \frac{q^2}{Q^6} - \frac{4(D-2)}{D-1} \frac{q^4}{Q^8} \bigg]\bigg\} \,,
\eqa
where
\beq\label{formulasoft2}
\int_{\textbf{p}}\frac{1}{p^2+m^2} =  - \frac{m}{4\pi} \bigg( \frac{\mu}{2 m} \bigg)^{2\epsilon} \big[ 1+2\epsilon \big]\,,
\eeq
Setting $D=4$, using Eqs.~(\ref{formb0bn}) and (\ref{formulasoft2}), and the formulas in App.~\ref{IntegralHTLpt}, \eqref{ghsRDR} reduces to
\bqa\label{ghsRDR1}
&& \mathcal{F}_{3g+4g+gh}^{(hs),\textrm{RDR}} = -\frac{1}{24\pi}m_D T^3-\frac{11}{384\pi^3}\bigg(\frac{1}{\epsilon}+2\gamma+\frac{68}{33}    \bigg)  \nonumber \\ && \hspace{1cm} \times\bigg( \frac{\mu}{4\pi T}   \bigg)^{2\epsilon}  \times \bigg( \frac{\mu}{2m_D}  \bigg)^{2\epsilon}m_D^3T  \,.
\eqa

The form of the $(hs)$ contribution to $\mathcal{F}_{3qg}$ and $\mathcal{F}_{4qg}$ expanded to first order in $m_D^2$ can be obtained from Ref.~\cite{Andersen:2003zk}, and in the RDR scheme can be expressed as
\bqa\label{qhsRDR}
&& \mathcal{F}_{3qg+4qg}^{(hs),\textrm{RDR}} = T \int_{\textbf{p}} \frac{1}{p^2+m_D^2} \bigg[ 4f_1- \sumint_{\{Q\}} \frac{8q^2}{Q^4}\bigg]  + 4m_D^2 T \nonumber \\ && \times \int_{\textbf{p}} \frac{1}{p^2+m_D^2} \bigg\{ f_2 +\sumint_{\{Q\}}\bigg[ -\frac{2(2+D)}{D-1}\frac{q^2}{Q^6}  + \frac{8}{D-1} \nonumber \\ && \times \frac{q^4}{Q^8} \bigg]\bigg\}  -8 m_q^2 T \int_{\textbf{p}} \frac{1}{p^2+m_D^2} \bigg[ 3f_2 -\sumint_{\{Q\}}\frac{4q^2}{Q^6}  \bigg]\,.
\eqa
Setting $D=4$, using Eqs.~(\ref{formb0bn}), (\ref{forlumaf1}), and (\ref{formulasoft2}) together with the formulas in App.~\ref{IntegralHTLpt}, \eqref{qhsRDR} reduces to
\bqa\label{qhsRDR1}
\mathcal{F}_{3qg+4qg}^{(hs),\textrm{RDR}} &=& -\frac{1}{12\pi}m_DT^3 + \frac{1}{4\pi^3}m_D m_q^2T \nonumber \\ && \hspace{-0.55cm} +\frac{1}{48\pi^3}   \bigg(\frac{1}{\epsilon}+2\gamma+  \frac{4}{3} +4\log2  \bigg) \nonumber \\ &&   \times \bigg( \frac{\mu}{4\pi T}   \bigg)^{2\epsilon}  \bigg( \frac{\mu}{2m_D}  \bigg)^{2\epsilon}m_D^3 T\,.
\eqa

The $(hs)$ contribution to $\mathcal{F}_{4s}$, $\mathcal{F}_{3gs}$, and $\mathcal{F}_{4gs}$ were presented in \cite{Du:2020odw} and in the RDR scheme can be expressed as
\bqa\label{f4shsRDR}
&& \mathcal{F}_{4s+3gs+4gs}^{(hs),\textrm{RDR}}= T\int_{\textbf{p}}\frac{1}{p^2+M_D^2}\bigg[ 3(D+4) b_1 -3M_D^2b_2  \nonumber \\ && \hspace{0.2cm} -3m_D^2b_2 - M_D^2 \sumint_{Q}\frac{12q^2}{(D-1)Q^6} \bigg] + T\int_{\textbf{p}}\frac{1}{p^2+m_D^2} \nonumber \\ && \times \bigg\{ -3 b_1  + 9M_D^2b_2  -6 m_D^2 b_2 + \sumint_{Q}\bigg[ \frac{6q^2}{Q^4}-M_D^2\frac{12q^2}{Q^6} \nonumber \\ && \hspace{1cm} + m_D^2  \bigg( 6\frac{D+3}{D-1}\frac{q^2}{Q^6} -\frac{24q^4}{(D-1)Q^8}\bigg)\bigg]  \bigg\},
\eqa
Setting $D=4$, using Eq.~(\ref{formb0bn}), and the formulas in App.~\ref{IntegralHTLpt}, \eqref{f4shsRDR} reduces to
\bqa\label{f4shsRDR1}
&& \mathcal{F}_{4s+3gs+4gs}^{(hs),\textrm{RDR}}= -T^3\bigg(\frac{m_D}{8\pi}+ \frac{M_D}{2\pi}  \bigg)-\frac{3}{32\pi^3}M_D^2 m_DT\nonumber \\
&&+\frac{3}{64\pi^3}\bigg( \frac{1}{\epsilon}+2+2\gamma \bigg)\bigg( \frac{\mu}{4\pi T}   \bigg)^{2\epsilon}\bigg( \frac{\mu}{2M_D}  \bigg)^{2\epsilon} M_D m_D^2T  \nonumber \\
&& +\frac{3}{32\pi^3} \bigg( \frac{1}{\epsilon}+\frac{5}{3}+2\gamma \bigg)\bigg( \frac{\mu}{4\pi T}   \bigg)^{2\epsilon}\bigg( \frac{\mu}{2M_D}  \bigg)^{2\epsilon} M_D^3T  \nonumber \\
&& +\frac{1}{128\pi^3} \bigg(\frac{1}{\epsilon}+\frac{16}{3}+2\gamma \bigg)\bigg( \frac{\mu}{4\pi T}   \bigg)^{2\epsilon}\bigg( \frac{\mu}{2m_D}  \bigg)^{2\epsilon}m_D^3 T\, . \nonumber \\
\eqa

Finally, the $(hs)$ contribution to $\mathcal{F}_{3qs}$ was first presented in \cite{Du:2020odw} and in the RDR scheme is
\bqa\label{fqshsRDR}
&& \mathcal{F}_{3qs}^{(hs),\textrm{RDR}} = 24T\int_{\textbf{p}}\frac{1}{p^2+M_D^2}\bigg[ -f_1 - M_D^2 f_2 \nonumber \\ &&\hspace{1cm} + 2m_q^2f_2  + M_D^2\sumint_{\{Q\}} \frac{2q^2}{(D-1)Q^6}    \bigg] ,
\eqa
Setting $D=4$, using Eqs.~(\ref{formb0bn}), (\ref{forlumaf1}), and (\ref{formulasoft2}) together with the formulas in App.~\ref{IntegralHTLpt}, \eqref{fqshsRDR} reduces to
\bqa\label{fqshsRDR1}
&& \mathcal{F}_{3qs}^{(hs),\textrm{RDR}} = -\frac{1}{4\pi}M_DT^3 +\frac{3}{16\pi^3} \bigg( \frac{1}{\epsilon}+\frac{8}{3}+2\gamma+4\log2  \bigg) \nonumber \\ && \hspace{0.6cm}\times \bigg( \frac{\mu}{4\pi T}   \bigg)^{2\epsilon}\bigg( \frac{\mu}{2M_D}  \bigg)^{2\epsilon} M_D^3T  -\frac{3}{4\pi^3}  \bigg( \frac{1}{\epsilon}+2+2\gamma \nonumber \\ && \hspace{1.2cm}+ 4\log2  \bigg)  \bigg( \frac{\mu}{4\pi T}   \bigg)^{2\epsilon}\bigg( \frac{\mu}{2M_D}  \bigg)^{2\epsilon}M_D m_q^2T\,.
\eqa

\section{NLO HTL thermodynamic potential in the RDR scheme}
\label{chap0402}

In this section, we will combine all contributions and counterterms to obtain the LO and NLO HTLpt-resummed $\symff$ thermodynamic potential $\Omega^{\textrm{RDR}}(T, \lambda, m_D, M_D, m_q, \delta=1)$. In the following subsections, we list the various terms contributing and compare them to the corresponding results obtained in the DRG scheme.

\subsection{Leading order}\label{leadingRDR}

By combining Eq.~(\ref{ghRDR1}) and contributions which are the same as in DRG scheme from Ref.~\cite{Du:2020odw}, our final result for the one-loop free energy is
\bqa\label{1opRDR}
&& \Omega_{\textrm{1-loop}}^{\textrm{RDR}} =  \mathcal{F}_{\textrm{ideal}}\bigg\{1-\hat{m}_D^2+4\hat{m}_D^3-6\hat{M}_D^2+24\hat{M}_D^3 \nonumber \\
&&\hspace{0.2cm} -8\hat{m}_q^2+16\hat{m}_q^4(6-\pi^2) +\frac{3}{4}\hat{m}_D^4\bigg[\frac{1}{\epsilon}+2\gamma+\frac{2\pi^2}{3} -\frac{16}{3} \nonumber \\
&&\hspace{0.2cm} + 2\log\frac{\hat{\mu}}{2}  -\frac{8\log2}{3}  \bigg]   +9\hat{M}_D^4\bigg[\frac{1}{\epsilon}+ 2\gamma+2\log\frac{\hat{\mu}}{2} \bigg]   \bigg\} \, , 
\eqa
where $\hat{m}_D$, $\hat{M}_D$, $\hat{m_q}$ and $\hat{\mu}$ are dimensionless variables, which are defined as
\bqa\label{dv}
 \hat{m}_D&=&\frac{m_D}{2\pi T} \, ,  \qquad
 \hat{M}_D=\frac{M_D}{2\pi T} \, , \nonumber \\
 \hat{m}_q&=&\frac{m_q}{2\pi T} \, ,   \qquad
 \hat{\mu}=\frac{\mu}{2\pi T}  \, .
\eqa 
Comparing to the one-loop free energy $\Omega_{\textrm{1-loop}}$ obtained in the DRG scheme, the only difference are the finite contributions which are proportional to $m_D^4$. Since the RDR divergences are the same as in the DRG scheme, the corresponding leading order vacuum energy counterterm $\Delta_0 \mathcal{E}_0$ is also the same. After adding $\Delta_0 \mathcal{E}_0$ to (\ref{1opRDR}), our final result for the LO renormalized thermodynamic potential in the RDR scheme is
\bqa\label{1oopRDR}
&& \frac{\Omega_{\textrm{LO}}^{\textrm{RDR}}}{\mathcal{F}_{\textrm{ideal}}} =  1-\hat{m}_D^2+4\hat{m}_D^3-6\hat{M}_D^2+24\hat{M}_D^3 -8\hat{m}_q^2 \nonumber \\
&&  \hspace{3mm}+ 16\hat{m}_q^4(6-\pi^2) +\frac{3}{4}\hat{m}_D^4\bigg[-\frac{16}{3} -\frac{8\log2}{3} + 2\gamma \nonumber \\
&&\hspace{1.3cm} +\frac{2\pi^2}{3}  +2\log\frac{\hat{\mu}}{2}    \bigg]   +18\hat{M}_D^4\bigg[ \gamma+\log\frac{\hat{\mu}}{2} \bigg]      \,.
\eqa

\subsection{Next-to-leading order}

By combining Eqs.~(\ref{ghhRDR}), (\ref{qhhRDR}), (\ref{f4shhRDR1}),  (\ref{ghsRDR}), (\ref{qhsRDR}), (\ref{f4shsRDR1}), (\ref{fqshsRDR1}), and contributions which are the same as in DRG scheme from Ref.~\cite{Du:2020odw}, and multiplying by $\lambda d_A$, we obtain the final result for the two-loop HTLpt thermodynamic potential
\begin{widetext}
\bqa\label{2opRDR}
&& \frac{\Omega_{\textrm{2-loop}}^{\textrm{RDR}}}{\mathcal{F}_{\textrm{ideal}}} =  \frac{\lambda}{\pi^2} \bigg\{-\frac{3}{2}+3\hat{m}_D+9\hat{M}_D-\frac{9}{2}\hat{M}_D\hat{m}_D  +\frac{9}{2}\hat{m}_D \hat{M}_D^2  -12\hat{m}_D \hat{m}_q^2 +\frac{3}{8} \hat{m}_D^2 \bigg[\frac{1}{\epsilon}+ 4\log\hat{m}_D \nonumber \\
&& \hspace{0.5cm} +4\log\frac{\hat{\mu}}{2}  + 6.32087357  \bigg] + \frac{9}{4} \hat{M}_D^2\bigg[ \frac{1}{\epsilon}+ 4\log\hat{M}_D+4\log\frac{\hat{\mu}}{2}   +5.32488132         \bigg]-\hat{m}_D^3\bigg[\frac{1}{2}+4\log2   \bigg] \nonumber \\
&& \hspace{1.5cm} -6\hat{m}_q^2 \bigg[\frac{1}{\epsilon}+4\log\frac{\hat{\mu}}{2}   +9.967511121   \bigg]- \hat{M}_D^3\bigg(36\log2+\frac{9}{2} \bigg)  +144\log2\hat{M}_D\hat{m}_q^2  \nonumber \\
&& \hspace{1.6cm} +\bigg[-\frac{27}{2} \hat{M}_D^3-\frac{9}{4} \hat{M}_D\hat{m}_D^2+36\hat{M}_D\hat{m}_q^2\bigg]  \bigg[2+\frac{1}{\epsilon}+2\gamma-2\log\hat{M}_D+4\log\frac{\hat{\mu}}{2}\bigg]
    \bigg\} . \hspace{7mm}
\eqa
\end{widetext}
Comparing to the two-loop DRG result $\Omega_{\textrm{2-loop}}$, we find that there are differences in the finite contributions proportional to $m_D^2$, $M_D^2$, $m_D^3$, $m_q^2$ and $M_D^3$.

The NLO RDR HTLpt counterterm contribution is obtained from the sum of Eqs.~(\ref{gcthRDR}) and contributions which are the same as in DRG scheme from Ref.~\cite{Du:2020odw}, giving
\bqa\label{cou2RDR}
&& \frac{\Omega_{\textrm{HTL}}^{\textrm{RDR}}}{\mathcal{F}_{\textrm{ideal}}} =  \hat{m}_D^2-6\hat{m}_D^3+6\hat{M}_D^2-36\hat{M}_D^3+8\hat{m}_q^2 \nonumber \\
&&\hspace{0.5cm} +32\hat{m}_q^4(\pi^2-6) -\frac{3}{2}\hat{m}_D^4\bigg[\frac{1}{\epsilon}+2\gamma+2\log\frac{\hat{\mu}}{2}+\frac{2\pi^2}{3} \nonumber \\
&&\hspace{0.8cm} -\frac{16}{3}- \frac{8\log2}{3} \bigg] -18\hat{M}_D^4\bigg[\frac{1}{\epsilon}+2\gamma+2\log\frac{\hat{\mu}}{2}\bigg]  . \hspace{7mm}
\eqa
Comparing to the DRG scheme result for $\Omega_{\textrm{HTL}}$, the only difference is the finite contribution proportional to $m_D^4$. We find that the form of $\Delta_1 \mathcal{E}_0$ is the same as in the DRG scheme. Since the divergent terms in Eq.~(\ref{2opRDR}) are the same as the ones in the DRG scheme, we find the same result for $\Delta\Omega_{\rm NLO}^{\textrm{RDR}}$ as in the DRG scheme.

By adding the one-loop (\ref{1oopRDR}), and two-loop (\ref{2opRDR}) HTLpt thermodynamic potentials, the HTLpt gluon and quark counterterms, and the HTLpt vacuum and mass renormalizations, we obtain the final expression for the NLO HTL thermodynamic potential in $\symff$ theory in the RDR scheme
\begin{widetext}
\bqa\label{loo2pRDR}
&& \frac{\Omega_{\textrm{NLO}}^{\textrm{RDR}}}{\mathcal{F}_{\textrm{ideal}}} =  1 -2\hat{m}_D^3-12\hat{M}_D^3+16\hat{m}_q^4(\pi^2-6)  -18\hat{M}_D^4\bigg(\gamma+\log\frac{\hat{\mu}}{2} \bigg)   -\frac{3}{2}\hat{m}_D^4\bigg( \gamma+\frac{\pi^2}{3}+\log\frac{\hat{\mu}}{2} -\frac{8}{3}  -\frac{4\log 2}{3} \bigg) \nonumber \\
&& \hspace{1.2cm} +\frac{\lambda}{\pi^2}\bigg[-\frac{3}{2}+3\hat{m}_D+9\hat{M}_D -\frac{9}{2}\hat{m}_D\hat{M}_D  +\frac{9}{2}\hat{m}_D \hat{M}_D^2-12\hat{m}_D \hat{m}_q^2 -12\hat{m}_q^2 \bigg(1.99870184+\log\frac{\hat{\mu}}{2}     \bigg)\nonumber \\
&& +\frac{3}{4} \hat{m}_D^2 \bigg(0.1753830597 + 2\log\hat{m}_D+\log\frac{\hat{\mu}}{2}   \bigg)  -\hat{m}_D^3\bigg(\frac{1}{2}+4\log2   \bigg)  +\frac{9}{2}\hat{M}_D^2\bigg( -0.3226130662  + 2\log\hat{M}_D+\log\frac{\hat{\mu}}{2}     \bigg) \nonumber \\
&&\hspace{1cm}  -\frac{9}{2} \hat{M}_D\hat{m}_D^2\bigg(\gamma+\log\frac{\hat{\mu}}{2}\bigg)  + 72\hat{M}_D\hat{m}_q^2\bigg(\gamma+2\log2+\log\frac{\hat{\mu}}{2} \bigg) - 9\hat{M}_D^3 \bigg(\frac{1}{2}+ 3\gamma+4\log2+3\log\frac{\hat{\mu}}{2}  \bigg)  \bigg]  \,.
\eqa
\end{widetext}
Note that the final expression is free from singularities and valid for all $N_c$.
Comparing to the result obtained using the DRG scheme in Ref.~\cite{Du:2020odw}, we find differences in terms without logarithms, which are proportional to $m_D^4$, $\lambda m_q^2$, $\lambda m_D^2$, $\lambda M_D^2$, $\lambda m_D^3$ and $\lambda M_D^3$.  These terms all contribute to the coefficient of $\lambda^2$ in the strict perturbative limit.

\subsection{The strict perturbative limit}

We next consider the strict perturbative limit through order $\lambda^2$.  To obtain this limit we evaluate the NLO HTLpt thermodynamic potential with LO perturbative gluon, scalar, and quark masses.
Expressed numerically, the NLO HTLpt result in the DRG scheme obtained in Ref.~\cite{Du:2020odw} is
\bqa\label{DRGHTLnumerical}
&& \frac{\Omega_{\textrm{NLO}}^{\textrm{DRG}}}{\mathcal{F}_{\textrm{ideal}}}\bigg|_{\hat{M}_{D} =\hat{M}_{D,\textrm{LO}}}^{\substack{ \hat{m}_{q}=\hat{m}_{q,\textrm{LO}}  \\ \hat{m}_{D} = \hat{m}_{D,\textrm{LO}}}}  =  1 - 0.151982 ~\lambda + 0.142365 ~\lambda^{3/2} \nonumber \\
&& \hspace{1.8cm} - 
 0.0923192 ~\lambda^2  + 
 0.015399 ~\lambda^2  \log\frac{\lambda}{\hat\mu}   \, ,
\eqa
where $\hat{m}_{q,\textrm{LO}}$, $\hat{m}_{D,\textrm{LO}}$, and $\hat{M}_{D,\textrm{LO}}$ are taken to be given by their leading-order weak-coupling limits given in Eq.~(\ref{mass}). The corresponding RDR HTLpt result obtained herein is 
\bqa\label{RDRHTLnumerical}
&& \frac{\Omega_{\textrm{NLO}}^{\textrm{RDR}}}{\mathcal{F}_{\textrm{ideal}}}\bigg|_{\hat{M}_{D} =\hat{M}_{D,\textrm{LO}}}^{\substack{ \hat{m}_{q}=\hat{m}_{q,\textrm{LO}}  \\ \hat{m}_{D} = \hat{m}_{D,\textrm{LO}}}}  =  1 - 0.151982 ~\lambda + 0.142365 ~\lambda^{3/2} \nonumber \\
&&\hspace{1.8cm} - 0.0912209 ~\lambda^2   + 0.015399 ~\lambda^2  \log\frac{\lambda}{\hat\mu} .
\eqa
We find the same dependence on the renormalization scale $\hat{\mu}$ and all terms are same except the term proportional to $\lambda^2$ without logarithms.  We note that the coefficient of $\lambda^2 \log \lambda$ is the same using both schemes, showing that it is scheme independent.

As further a comparison, the perturbative free energy computed to three-loop order, keeping all terms through order $\lambda^2$ was obtained in Refs.~\cite{Du:2021jai,Andersen:2021bgw}.  Expressed numerically the full result is
\bqa\label{RDRsec3numerical}
\frac{F_\text{3-loop}^{\textrm{resum}}}{\mathcal{F}_{\textrm{ideal}}} &=& 1 - 0.151982 ~\lambda + 0.142365 ~\lambda^{3/2} \nonumber \\
&& \hspace{2mm} -0.0613173 ~\lambda^2 + 0.015399 ~\lambda^2 \log\lambda \,.
\eqa

Comparing Eqs.~(\ref{DRGHTLnumerical}) and (\ref{RDRHTLnumerical}) to (\ref{RDRsec3numerical}), we find all terms are the same except the terms proportional to $\lambda^2$.  This is expected since both the RDR and DRG scheme HTLpt calculations were two-loop calculations and hence cannot reproduce the full $\lambda^2$ coefficient.
At two-loop order one could consider fixing $\hat\mu$ as to reproduce the full order $\lambda^2$ coefficient.  This occurs when $\hat{\mu} \simeq 0.133554$ and $\hat{\mu} \simeq 0.143428 $, in the DRG and RDR schemes, respectively.
\subsection{Gap equations}

To go beyond strict perturbation theory we need to fix the mass parameters in a non-perturbative manner. Using the NLO HTLpt thermodynamic potential, the gluon, scalar, and quark masses can be fixed using the variational method described previously. The gluon, scalar, and quark mass parameters $m_D$, $M_D$, and $m_q$ can be determined by requiring the derivative of $\Omega_{\textrm{NLO}}^{\textrm{RDR}}$ with respect to each parameter is zero.
\bqa\label{gap1RDR}
\frac{\partial}{\partial m_q}\Omega_{\textrm{NLO}}^{\textrm{RDR}}(T, \lambda, m_D, M_D, m_q, \delta=1)&=&0 \, ,    \nonumber \\
\frac{\partial}{\partial m_D}\Omega_{\textrm{NLO}}^{\textrm{RDR}}(T, \lambda, m_D, M_D, m_q, \delta=1)&=&0 \, ,    \nonumber \\
\frac{\partial}{\partial M_D}\Omega_{\textrm{NLO}}^{\textrm{RDR}}(T, \lambda, m_D, M_D, m_q, \delta=1)&=&0    \,.
\eqa
The first equation in Eq.~(\ref{gap1RDR}) gives
\bqa\label{gapmqRDR}
&& \hat{m}_q^2\big(\pi^2-6 \big) = \frac{\lambda}{4\pi^2}\bigg[\frac{3}{2}\hat{m}_D+\frac{3}{2} \bigg(1.998701836+\log\frac{\hat{\mu}}{2}   \bigg) \nonumber \\ && \hspace{2cm} - 9 \hat{M}_D\bigg( \gamma+2\log2+\log\frac{\hat{\mu}}{2} \bigg) \bigg] .
\eqa
The second equation gives
\bqa\label{gapmdRDR}
&&\hspace{-2mm}\hat{m}_D^2+\hat{m}_D^3\bigg(\!\gamma+\frac{\pi^2}{3}+ \log\frac{\hat{\mu}}{2} - \frac{8}{3} -\frac{4\log2}{3}\bigg) \!=\! \frac{\lambda}{4\pi^2}\bigg[2 \nonumber \\
&&\hspace{0.5cm} -3\hat{M}_D  +3\hat{M}_D^2  -8\hat{m}_q^2  -\hat{m}_D^2\bigg(\!1+8\log2   \bigg)     \nonumber \\
&&\hspace{1cm} -6\hat{m}_D \hat{M}_D\bigg(\gamma+ \log\frac{\hat{\mu}}{2}  \bigg)   \nonumber \\
&& \hspace{0.5cm}+\hat{m}_D\bigg( 1.1753831  +2\log\hat{m}_D+ \log\frac{\hat{\mu}}{2}   \bigg)\bigg] .
\eqa
Finally, the third equation gives
\bqa\label{gapmRDR}
&&\hat{M}_D^2+2\hat{M}_D^3\bigg( \gamma+ \log\frac{\hat{\mu}}{2} \bigg)= \frac{\lambda}{4\pi^2}\bigg[1-\frac{1}{2}\hat{m}_D +\hat{m}_D \hat{M}_D \nonumber \\
&&\hspace{0.3cm} -\frac{1}{2}\hat{m}_D^2\bigg(\gamma+ \log\frac{\hat{\mu}}{2}  \bigg)  + 8\hat{m}_q^2\bigg(\gamma+ 2\log2+ \log\frac{\hat{\mu}}{2}  \bigg)  \nonumber \\
&&\hspace{1cm} - 3\hat{M}_D^2\bigg(4\log2+3\gamma +3\log\frac{\hat{\mu}}{2} + \frac{1}{2}\bigg)\nonumber \\
&&\hspace{0.5cm}   + \hat{M}_D \bigg(0.677386934 +2 \log\hat{M}_D+\log\frac{\hat{\mu}}{2} \bigg)         \bigg] .
\eqa

We find that, similar to the DRG scheme, $\hat{m}_q$ in Eqs.~(\ref{gapmdRDR}) and (\ref{gapmRDR}) can be written in terms of $\hat{M}_D$ and $\hat{m}_D$ by using (\ref{gapmqRDR}). By numerically solving these three equations simultaneously, we obtain the gap equation solutions, $\hat{m}_q^{\textrm{gap}}(\lambda,\hat\mu)$, $\hat{m}_D^{\textrm{gap}}(\lambda,\hat\mu)$, and $\hat{M}_D^{\textrm{gap}}(\lambda,\hat\mu)$.

\begin{figure}[t]
\centering
\includegraphics[width=0.39\textwidth]{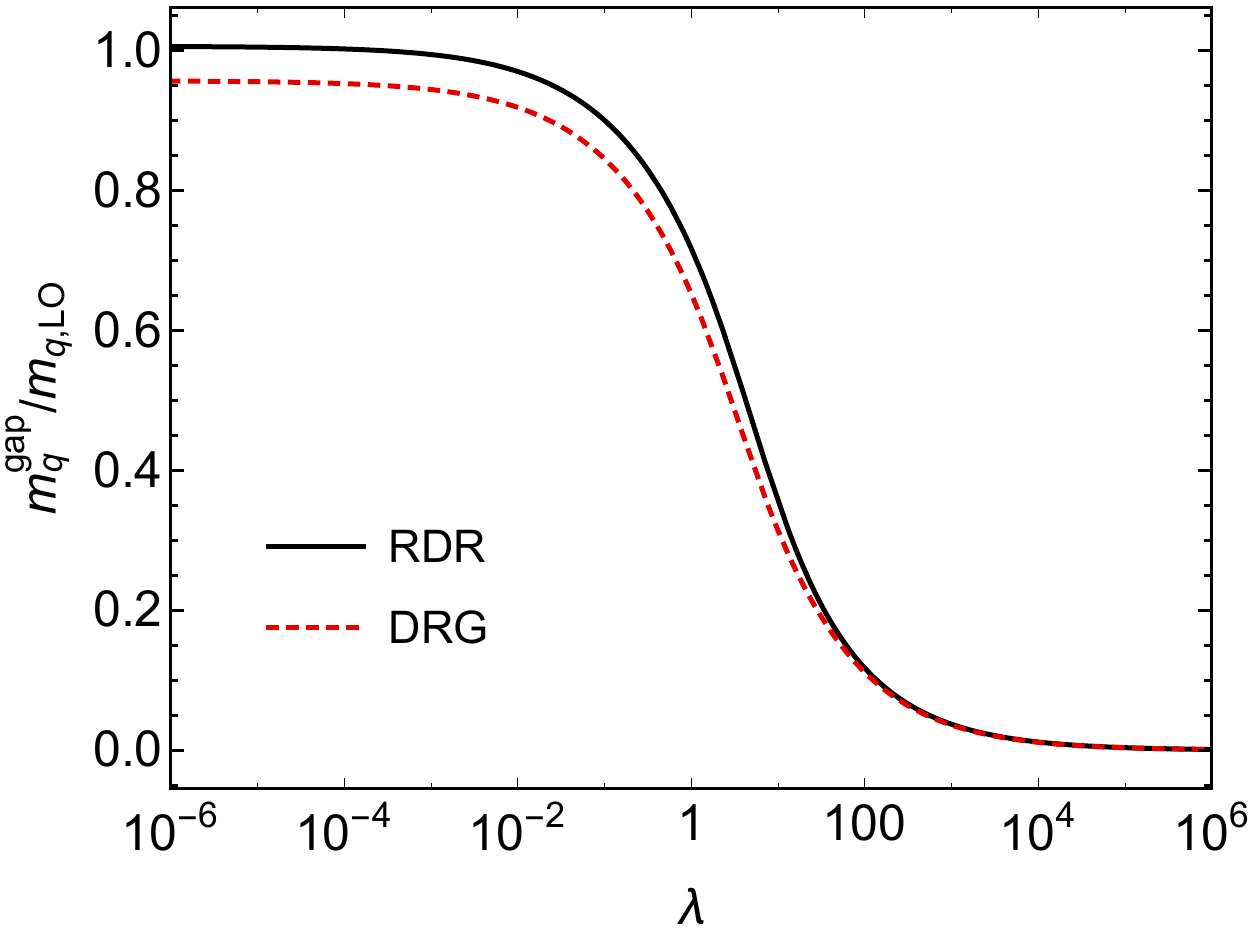}\\[2ex]
\includegraphics[width=0.39\textwidth]{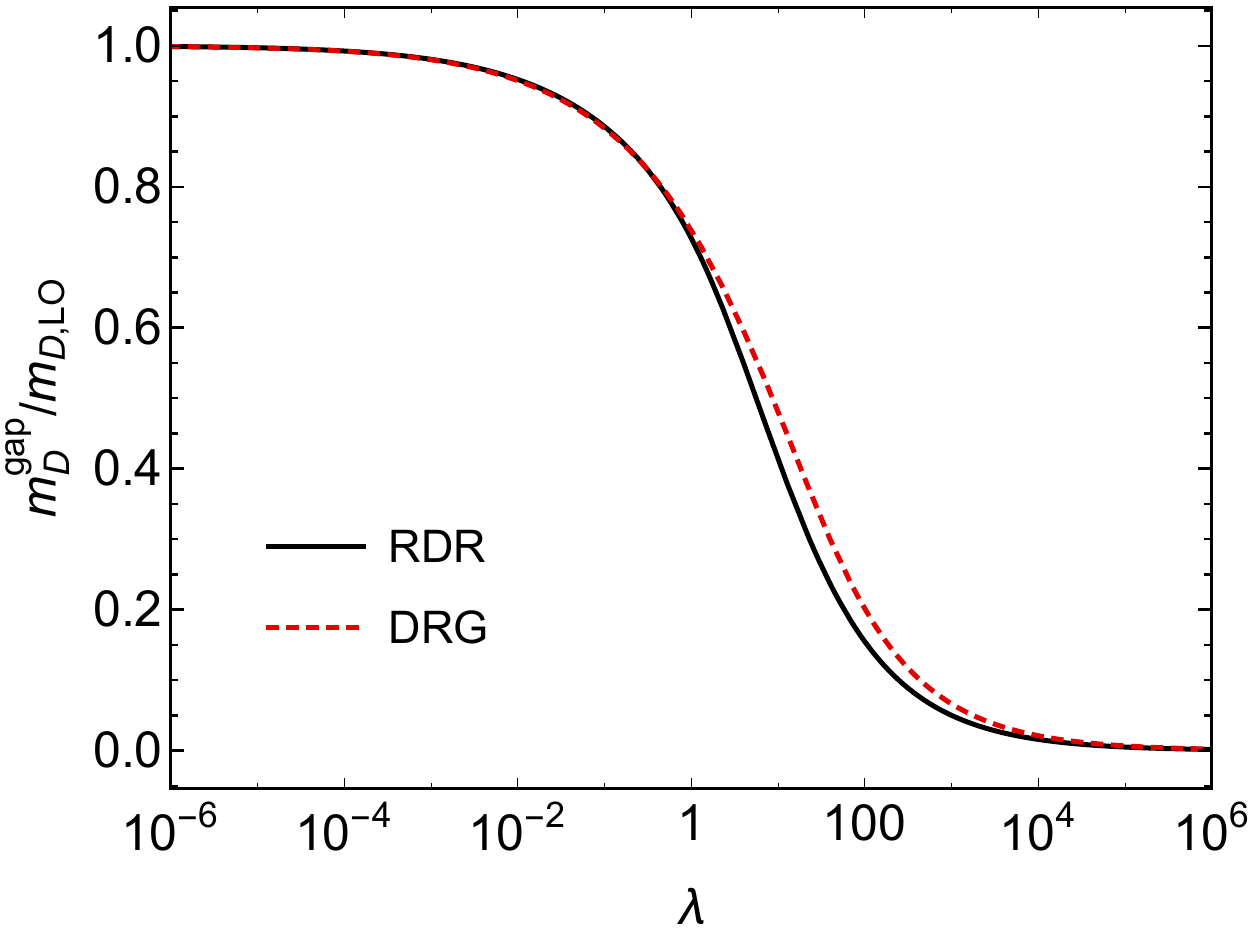}\\[2ex]
\includegraphics[width=0.39\textwidth]{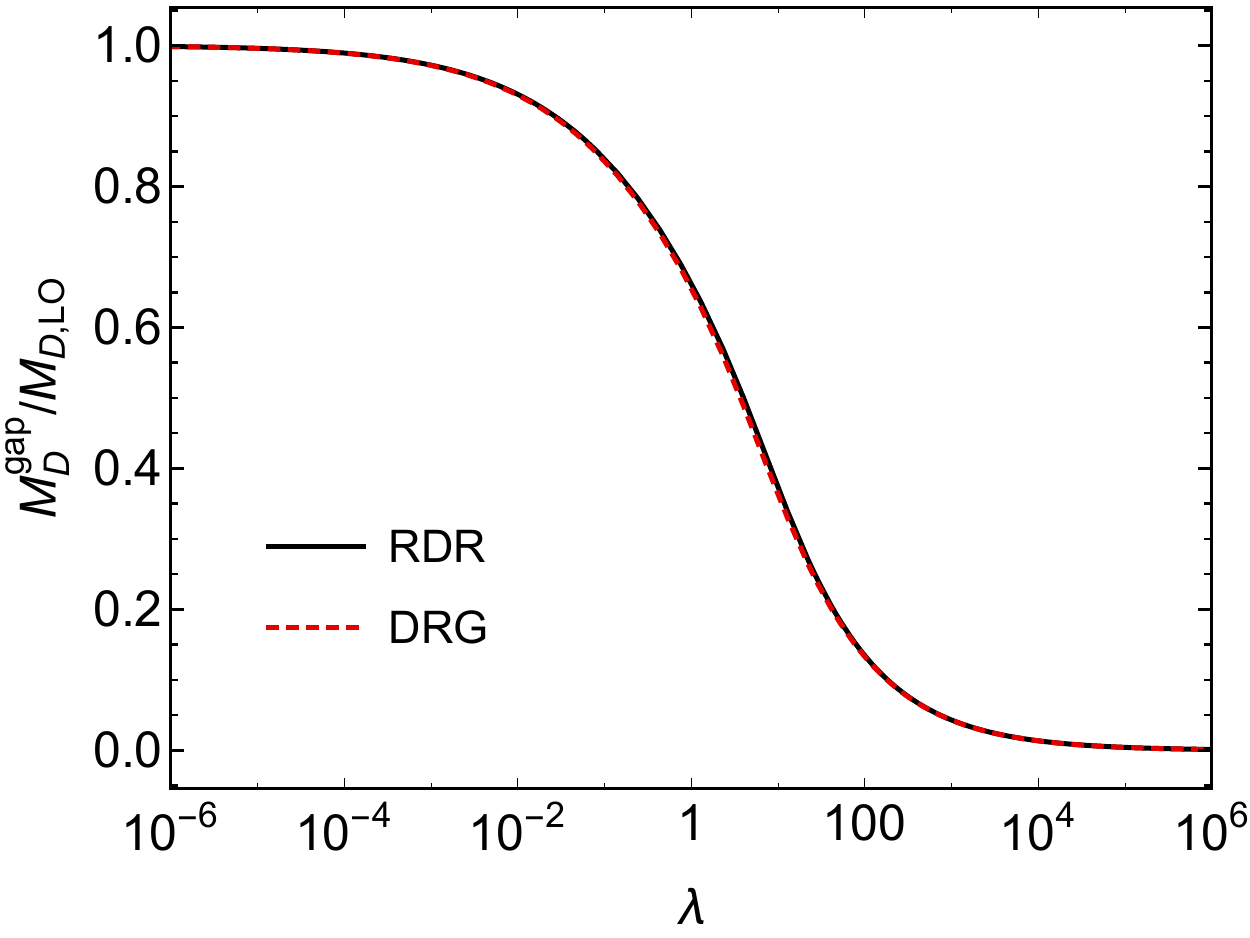}
\vspace*{-.08cm}
\caption{\label{RDRgapsol}
(Color online) Comparisons of the numerical solution of gap equations in the RDR and DRG schemes for $m_q$, $m_D$, and $M_D$ as a function of $\lambda$ with $\hat{\mu}=1$. In each panel, the results are scaled by their corresponding leading-order weak-coupling limits.}
\end{figure}

In Fig.~\ref{RDRgapsol}, we compare numerical solution of quark, gluon, and scalar gap equations in the RDR scheme to the result obtained using the DRG scheme. In these three panels, the results are obtained using the central value of the renormalization scale, $\hat{\mu}=1$, and are scaled by their corresponding leading-order weak-coupling limits. The red line is the result obtained using the RDR scheme, and the black dotted line is the result obtained using the DRG scheme. As one can see from Fig.~\ref{RDRgapsol}, the results obtained in these two schemes are numerically similar to each other. At the central value $\mu=2\pi T$, for the quark mass, the RDR result is slightly higher than the DRG result when $\lambda \rightarrow 0$;\footnote{The gap equation solution for $m_q$ does not approach its perturbative limit when $\lambda$ approaches to zero in both schemes.} for the gluon mass, the RDR result is a little lower than the DRG result at intermediate $\lambda$; for the scalar mass, the two results are nearly identical.

\section{Thermodynamic functions in the RDR scheme}
\label{chap0403}

The NLO HTLpt-resummed free energy in the RDR scheme can be obtained by evaluating the NLO thermodynamic potential (\ref{loo2pRDR}) at the solution of the gap equations (\ref{gap1RDR})
\beq
\mathcal{F}_{\rm NLO}^{\textrm{RDR}} = \Omega_{\rm NLO}^{\textrm{RDR}} (T, \lambda, m_D^{\rm gap},  M_D^{\rm gap}, 
m_q^{\rm gap}, \delta=1) \, .
\eeq
The pressure, entropy density, and energy density can be obtained using 
\bqa\label{PSE}
\mathcal{P} &=& -\mathcal{F} \, , \nonumber \\
\mathcal{S} &=& -\frac{d \mathcal{F}}{d T}  \, , \nonumber \\
\mathcal{E} &=& \mathcal{F} - T\frac{d \mathcal{F}}{d T} \, .
\eqa
In $\symff$ these three quantities satisfy
\bqa 
\frac{\mathcal{P}}{\mathcal{P}_{\rm ideal}} = \frac{\mathcal{S}}{\mathcal{S}_{\rm ideal}} = \frac{\mathcal{E}}{\mathcal{E}_{\rm ideal}}  \; ,
\eqa
since $\lambda$ is temperature independent. 

\subsection{Numerical results}\label{resultsRDR}

\begin{figure}[t!]
\centering
\includegraphics[width=0.5\textwidth]{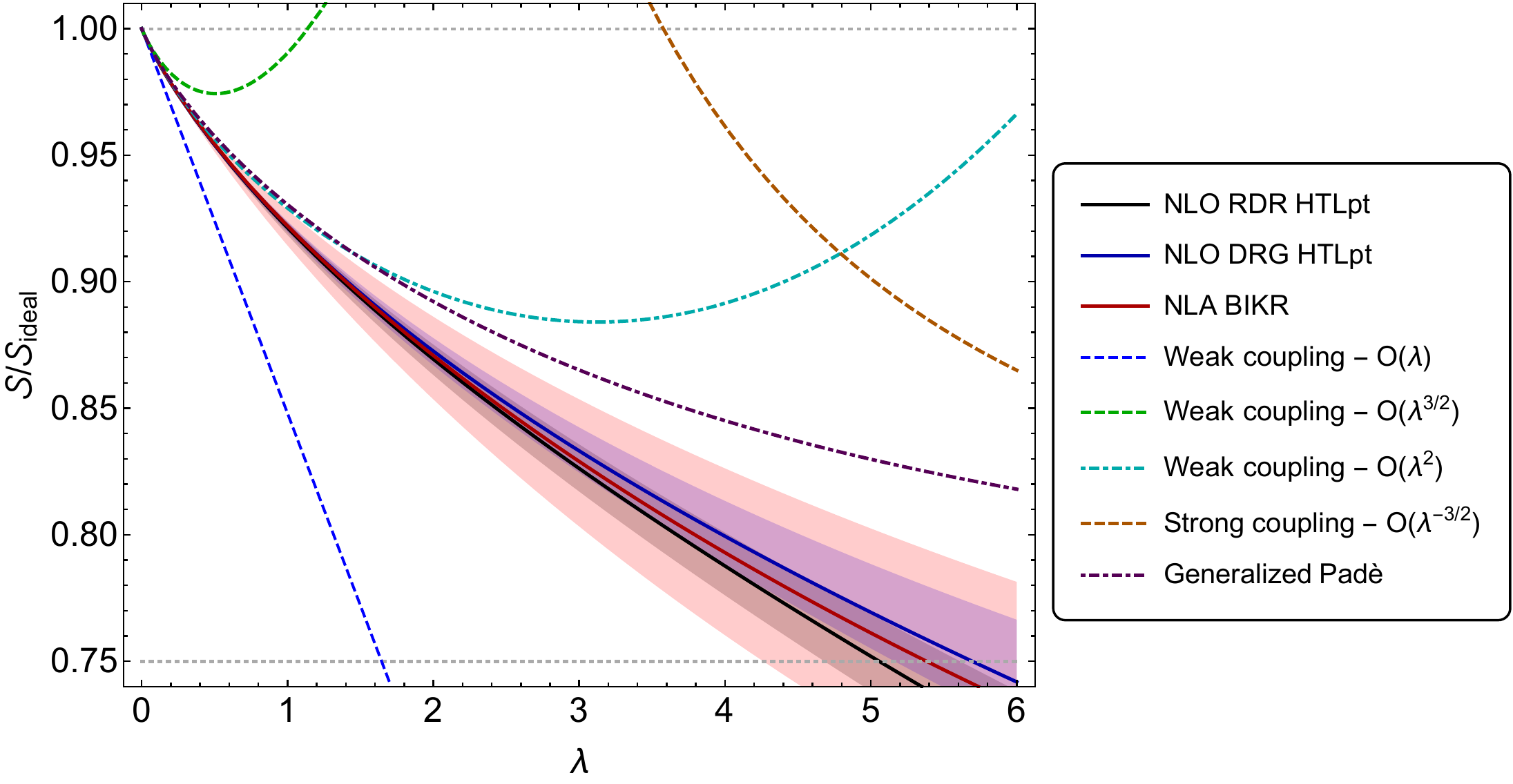}
\vspace*{-.08cm}
\caption{\label{HTLNLA1RDR}
(Color online) Comparison of the NLO scaled entropy density obtained in the RDR scheme, the DRG scheme, and the prior NLA work of Blaizot, Iancu, Kraemmer, and Rebhan (BIKR) \cite{Blaizot:2006tk}. A detailed description of the various lines can be found in the text.}
\end{figure}

\begin{figure}[t!]
\centering
\includegraphics[width=0.49\textwidth]{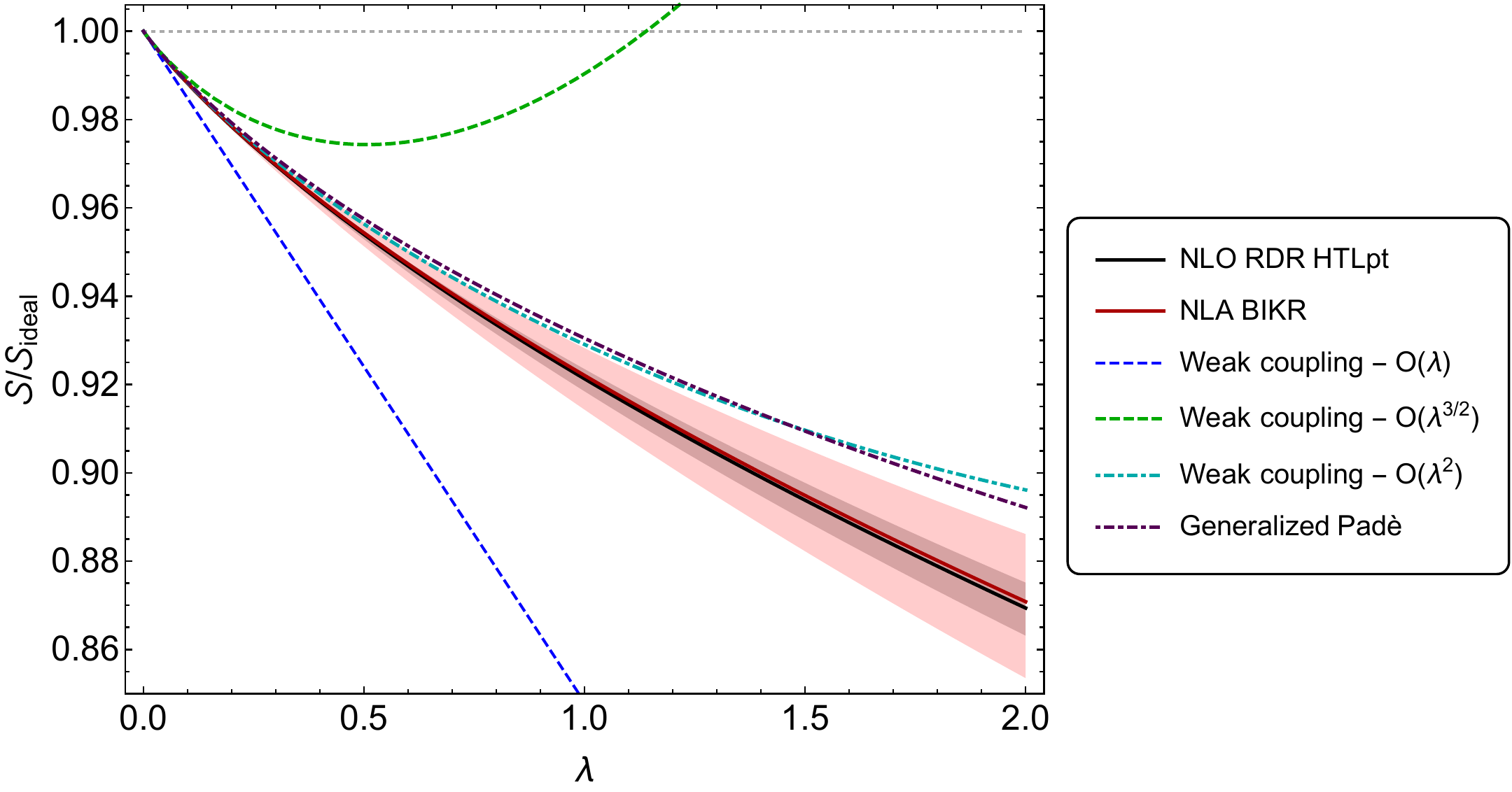}
\vspace*{-.08cm}
\caption{\label{HTLNLA2RDR}
(Color online) Comparison of our NLO result for the scaled entropy density in the RDR scheme with prior results for small $\lambda$.  Lines are the same as in fig.~\ref{HTLNLA1RDR}. }
\end{figure}

In Fig.~\ref{HTLNLA1RDR}, we present a comparison of our NLO HTLpt result in the RDR and DRG schemes with the next-to-leading approximation (NLA) result of Blaizot, Iancu, Kraemmer, and Rebhan (BIKR)~\cite{Blaizot:2006tk}.
The green, blue, and cyan dashed/dotted lines represent the strict perturbation theory result given in Eq.~\eqref{weaexp}, truncated at order $\lambda, \lambda^{3/2}, \lambda^2$ (including the logarithmic term), respectively. The orange dashed line is the strong coupling result given in Eq.~(\ref{stro}). The purple dot-dashed line is the result of a generalized Pad\'{e} approximant which is constructed based on Eqs.~(\ref{weaexp}) and (\ref{stro})~\cite{Andersen:2021bgw}. The black and blue lines with a shaded bands are the NLO HTLpt result in the RDR and DRG schemes, respectively. The bands correspond to variation of $\hat\mu$ in the range $1/2 \leq \hat\mu \leq 2$ and the solid lines correspond to $\hat\mu=1$.  Finally, the red line with a shaded band is the NLO BIKR result with the bands and line corresponding to scale variation and central value.  

As Fig.~\ref{HTLNLA1RDR} demonstrates, the DRG and RDR schemes result in similar predictions for $\lambda \lesssim 6$, however, both are below the generalized Pad\'{e} which is constructed using the full order $\lambda^2$ and $\lambda^2 \log \lambda$ coefficients.  Since the DRG scheme breaks supersymmetry, the RDR result should be taken as the correct NLO HTLpt result.  In comparison with the BIKR result, which made use of the DRG scheme, we see that it is also in agreement to within a few percent for $\lambda \lesssim 6$.

In Fig.~\ref{HTLNLA2RDR}, we present a similar comparison, but for smaller $\lambda$ in order to see the differences more clearly. The various strict perturbative results in this figure are the same as in Fig.~\ref{HTLNLA1RDR}. As can be seen from this figure, there is excellent agreement between the BIKR result and RDR NLO HTLpt in this coupling range. The generalized Pad\'{e} approximant, however, only overlaps with both resummed results for $\lambda \lesssim 0.5$.

\section{Conclusions}
\label{conclusions}

In this work, we obtained the LO and NLO HTLpt resummed thermodynamic potential in $\symff$ using the RDR regularization scheme.  We then analytically and numerically compared the RDR result to the result obtained in the DRG scheme. From the analytical perspective, we found that differences emerged only in finite contributions, while the divergences were unaffected. This causes only a small change compared to DRG HTLpt result, which can be seen from the figures presented in Sec.~\ref{resultsRDR}.  Importantly, we found that, when expanded in strict perturbation theory, both schemes gave the same results for the coefficients of all terms contributing at orders $\lambda$, $\lambda^{3/2}$, and $\lambda^2 \log\lambda$, while at order $\lambda^2$ we found a weak residual regularization scheme dependence.  

The results reported herein provide an important cross check of the coefficient of the order $\lambda^2 \log\lambda$ term in the free energy, independently confirming our prior result for this coefficient, which was obtained in Refs.~\cite{Du:2021jai,Andersen:2021bgw}.  The differences at order $\lambda^2$ are not yet conclusive, however, since in both the RDR and DRG schemes we have only performed a two-loop HTLpt calculation and a three-loop calculation is needed to properly fix this coefficient.  Such a calculation was performed in strict perturbation theory in Refs.~\cite{Du:2021jai,Andersen:2021bgw}.  It would be interesting to extend the RDR HTLpt calculation to three-loop order in order to see how well it performs at large coupling, however, we first plan to complete the computation of the order $\lambda^{5/2}$ coefficient using effective field theory methods since this calculation is somewhat more manageable.

\section*{Acknowledgements}

We thank Jens O. Andersen for discussions and comments.  Q.D. was supported by the Guangdong Major Project of Basic and Applied Basic Research No. 2020B0301030008 and the Natural Science Foundation of China Project No. 11935007. M.S. and U.T. were supported by the U.S. Department of Energy Award No.~DE-SC0013470.   

\appendix

\section{HTLpt Feynman rules for $\symff$ in the RDR scheme}\label{fmr}

In this appendix, we will present the RDR scheme HTLpt Feynman rules for $\symff$ in Minkowski space. In Minkowski space, the momentum is denoted by $p=(p_0,\textbf{p})$, and satisfies $p\cdot q =p_0 q_0-\textbf{p}\cdot \textbf{q}$. The four-vector $n^\mu$ specifies the thermal rest frame and in the local rest frame of the plasma given by $n=(1,\textbf{0})$.

We begin by noting that, since all ultraviolet divergences which are generated from the HTLpt reorganization are all canceled by systematically derivable counterterms, we only need to consider the leading order contributions to the gluon, quark, and scalar masses 
\beq\label{mass}
m_{D,\rm LO}^2 = 2\lambda T^2 \, , 
\quad
M_{D,\rm LO}^2 = \lambda T^2 \, ,  
\quad
m_{q,\rm LO}^2 = \frac{\lambda T^2}{2}  \, ,
\eeq
which have been given previously in Refs.~\cite{Du:2020odw}. 

\subsection{HTLpt gluon polarization tensor}\label{gsfn}

In $\symff$, for massless bosons and fermions, the gluon polarization tensor in HTL limit can be expressed as
\bqa\label{sfg}
\Pi^{\mu\nu}(p)=m_D^2\big[ \mathcal{T}^{\mu\nu}(p,-p)-n^\mu n^\nu   \big] \, ,
\eqa
where we have introduced a rank-two tensor $\mathcal{T}^{\mu\nu}(p,q)$, which is defined only when $p+q=0$,
\bqa\label{ts}
\mathcal{T}^{\mu\nu}(p,-p) \equiv \bigg\langle y^\mu y^\nu \frac{p \cdot n}{p \cdot y} \bigg \rangle_{\hat{\textbf{y}}}      \,.
\eqa
The angular brackets represent an average over the spatial direction of the light-like vector $y$. The tensor $\mathcal{T}^{\mu\nu}$ is symmetric in $\mu$ and $\nu$, and satisfies the ``Ward identity''
\bqa\label{wit}
p_\mu\mathcal{T}^{\mu\nu}(p,-p)=(p\cdot n) n^\nu    \,.
\eqa
As a result, the polarization tensor $\Pi^{\mu\nu}$ is also symmetric in  $\mu$ and $\nu$ and satisfies
\bqa\label{wis}
p_\mu \Pi^{\mu\nu}(p)&=& 0 \, ,   \nonumber \\  
g_{\mu\nu}\Pi^{\mu\nu}(p)&=&-m_D^2 \, .
\eqa

The gluon polarization tensor can also be expressed in terms of two scalar functions, the transverse and longitudinal polarization functions $\Pi_T$ and $\Pi_L$, which can be written as
\bqa\label{tslo0}
 \Pi_T(p)&=& \frac{1}{D-2}\big( \delta^{ij}-\hat{p}^i\hat{p}^j  \big) \Pi^{ij}(p) \, ,  \nonumber \\
 \Pi_L(p)&=& -\Pi^{00}(p)  \, ,
\eqa
where $\hat{\textbf{p}}=\textbf{p}/|\textbf{p}| $ is the unit vector in the direction of $\textbf{p}$. The gluon polarization tensor can be expressed in terms of these two functions
\bqa\label{sfg1}
\Pi^{\mu\nu}(p)=- \Pi_T(p)T_p^{\mu\nu} -\frac{1}{n_p^2}\Pi_L(p) L_p^{\mu\nu} \, ,
\eqa
with
\bqa\label{tl}
 T_p^{\mu\nu}&=& g^{\mu\nu}-\frac{p^\mu p^\nu}{p^2}- \frac{n_p^\mu n_p^\nu}{n_p^2} \, ,  \nonumber \\    L_p^{\mu\nu}&=& \frac{n_p^\mu n_p^\nu}{n_p^2} \, .
\eqa
The four-vector $n_p^\mu$ appearing above is defined by
\bqa\label{np}
 n_p^\mu=n^\mu- \frac{n\cdot p}{p^2}p^\mu\,,
\eqa
which satisfies $p\cdot n_p=0$ and $n_p^2=1-(n\cdot p)^2/p^2$. As a consequence, Eq.(\ref{wis}) reduces to
\bqa\label{wis1}
(D-2)\Pi_T(p)+\frac{1}{n_p^2}\Pi_L(p) =m_D^2 \, .
\eqa

In the HTL limit, the polarization functions $\Pi_T(p)$ and $\Pi_L(p)$ can be written in terms of $\mathcal{T}^{00}$
\bqa\label{tslo}
 \Pi_T(p)&=& \frac{m_D^2}{(D-2)n_p^2}\big[\mathcal{T}^{00}(p,-p)-1 +n_p^2 \big] \, ,  \nonumber \\  \Pi_L(p)&=& m_D^2\big[1-\mathcal{T}^{00}(p,-p)  \big]     \,.
\eqa
Note that, it is crucial to take the angular average in $d-1=3-2\epsilon$ in (\ref{ts}), and then analytically continue to $d-1=3$ only after all poles in $\epsilon$ have been cancelled. The expression for $\mathcal{T}^{00}$ is
\bqa\label{t00}
\hspace{-0.5cm}\mathcal{T}^{00}(p,-p)=\frac{\omega(\epsilon)}{2}\int^{1}_{-1}dc \, (1-c^2)^{-\epsilon} \frac{p_0}{p_0-|\textbf{p}|c} \, ,
\eqa
where the weight function $\omega(\epsilon)$ is given by
\bqa\label{t000}
 \omega(\epsilon)= \frac{\Gamma(2-2\epsilon)}{\Gamma^2(1-\epsilon)}2^{2\epsilon} =\frac{\Gamma( \frac{3}{2}  -\epsilon)}{\Gamma(\frac{3}{2})\Gamma(1-\epsilon)}    \,.
\eqa
In the imaginary-time formalism, $\mathcal{T}^{00}(p,-p)$ can be expressed as
\bqa\label{t00x1}
\mathcal{T}_P=\bigg\langle\frac{P_0^2}{P_0^2+p^2c^2}   \bigg\rangle_c \,,
\eqa
where the angular brackets represent an average over $c$ defined by
\bqa\label{t00x2}
\big\langle f(c)\big\rangle_c \equiv \omega(\epsilon) \int^1_0 dc (1-c^2)^{-\epsilon}f(c) \, .
\eqa

The definition of the gluon self energy in Eq.~(\ref{sfg}) is the same as in QCD in Ref.~\cite{Andersen:2002ey} up to the definition of $m_D$.  The HTLpt three-gluon vertex, four-gluon vertex, and ghost-gluon vertex are also similar C \cite{Du:2020odw}.

\subsection{HTLpt gluon propagator}

The HTLpt Feynman rule for the gluon propagator is
\bqa\label{prog}
i\delta^{ab}\Delta_{\mu\nu}(p)  \, ,
\eqa
where $a$ and $b$ are adjoint color indices and 
\bqa\label{prog7}
&& \Delta^{\mu\nu}(p)  = \big[-\Delta_T(p)g^{\mu\nu}+\Delta_X(p)n^\mu n^\nu \big] \nonumber \\ && \hspace{7mm} - \frac{n\cdot p}{p^2}\Delta_X(p) (p^\mu n^\nu+n^\mu p^\nu) \nonumber \\ && \hspace{7mm} + \bigg[ \Delta_T(p)+\frac{(n\cdot p)^2}{p^2}\Delta_X(p)-\frac{\xi}{p^2}  \bigg]\frac{p^\mu p^\nu}{p^2}\, ,
\eqa  
where $\Delta_T$ and $\Delta_L$ are the transverse and longitudinal propagators, respectively, and defined by
\bqa\label{prog2}
\Delta_T(p)&=&\frac{1}{p^2-\Pi_T(p)} \, ,   \nonumber \\  \Delta_L(p)&=&\frac{1}{-n_p^2p^2+\Pi_L(p)}   \,, 
\eqa
and $\Delta_X(p)$ is defined as 
\bqa\label{prog5}
 \Delta_X(p)=\Delta_L(p)+\frac{1}{n_p^2}\Delta_T(p)\,.
\eqa
%

\subsection{HTLpt quark self energy, propagator, and counterterm}
\label{qsfn}

The HTLpt quark self energy is
\bqa\label{sfq}
 \Sigma(p)=m_q^2{\displaystyle{\not} \mathcal{T}}(p)   \, ,
\eqa
where we have suppressed the trivial Kronecker deltas and
\bqa\label{sfq2}
 \mathcal{T}^\mu (p) \equiv \bigg\langle \frac{y^\mu}{p\cdot y} \bigg\rangle_{\hat{\textbf{y}}}\,,
\eqa

Similar to the gluon polarization tensor, the angular average in $ \mathcal{T}^\mu$ can be written as
\bqa\label{sfq3}
\mathcal{T}^\mu(p)=\frac{\omega(\epsilon)}{2}\int^{1}_{-1}dc(1-c^2)^{-\epsilon} \frac{y^\mu}{p_0-|\textbf{p}|c}      \,.
\eqa

As was the case with the gluonic self energy, the quark self energy (\ref{sfq}) is the same as in QCD \cite{Andersen:2003zk} up to the definition of $m_q$, and taking into account the four Majorana fermions which are indexed by $i$. This will also occur in the HTLpt quark propagator, quark-gluon three vertex, and quark-gluon four vertex in $\symff$ \cite{Du:2020odw}.

The HTLpt Feynman rule for the quark propagator is
\beq\label{proq}
i\delta^{ab}\delta^{ij}S(p) \, ,
\eeq
where $i,j$ index the four independent Majorana fermions and
\beq
S(p)=\frac{1}{{\displaystyle{\not} p}-\Sigma(p)} = \frac{1}{{\displaystyle{\not}\mathcal{A}}(p)} \, ,
\eeq
where $\mathcal{A}_\mu(p)=(\mathcal{A}_0(p),\mathcal{A}_s(p)\hat{\textbf{p}})$ with
\bqa\label{proq2}
\mathcal{A}_0(p)&=&p_0-\frac{m_q^2}{p_0}\mathcal{T}_p \, ,     \nonumber \\
\mathcal{A}_s(p)&=&|\textbf{p}|+\frac{m_q^2}{|\textbf{p}|} \big[1-\mathcal{T}_p  \big]   .
\eqa

The insertion of an HTL quark counterterm into a quark propagator corresponds to
\bqa\label{proq3}
i\delta^{ab}\delta^{ij}\Sigma(p) \, .
\eqa

\subsection{HTLpt  scalar self energy, propagator, and counterterm} \label{ssfn}

The HTL scalar self energy can be expressed as
\bqa\label{sfsy}
\mathcal{P}_{ab}^{AB}(p) =\delta_{ab} \delta^{AB} \lambda T^2  \, ,
\eqa 
and the corresponding Feynman rule for the scalar propagator is
\bqa\label{pros}
i \delta^{ab}\delta^{AB}\Delta_s(p) \,,
\eqa
where
\bqa\label{pros1}
\Delta_s(p)=\frac{1}{p^2-M_D^2} \,.
\eqa
The insertion of an HTL scalar counterterm into a scalar propagator corresponds to
\bqa\label{pros3}
-i \delta^{ab}\delta^{AB}\mathcal{P}_{aa}^{AA}(p)\,.
\eqa
%

\subsection{HTLpt quark-gluon vertex}
\label{subsec:qgv}

The quark-gluon vertex with incoming quark momentum $r$, incoming gluon momentum $p$, and outgoing quark momentum $q$, Lorentz index $\mu$, and color indices $b$, $a$, $c$ is
\bqa\label{3qg}
\Gamma_{abc}^{\mu,ij}(p,q,r)&=& -g f_{abc}\delta^{ij}\big[\gamma^\mu + m_q^2\tilde{\mathcal{T}}^\mu(p,q,r)  \big] \nonumber \\   &=& -g f_{abc}\delta^{ij}\Gamma^{\mu}(p,q,r) \, .
\eqa
The rank-one tensor $\tilde{\mathcal{T}}^\mu$ is only defined for $p+r-q=0$
\bqa\label{3qg2}
\tilde{\mathcal{T}}^\mu(p,q,r) \equiv \bigg\langle y^\mu\bigg(\frac{{\displaystyle{\not} y}}{(y\cdot r)(y\cdot q)}  \bigg)     \bigg\rangle_{\hat{\textbf{y}}} \, ,
\eqa
and is even under the permutation of $q$ and $r$. It satisfies
\bqa\label{3qg3}
p_\mu \tilde{\mathcal{T}}^\mu(p,q,r) ={\displaystyle{\not} \mathcal{T}} (r)-{\displaystyle{\not} \mathcal{T}}(q) \, .
\eqa
The quark-gluon vertex therefore satisfies the Ward identity
\bqa\label{3qg4}
p_\mu \Gamma^\mu(p,q,r)=S^{-1}(q)-S^{-1}(r) \,.
\eqa

\subsection{HTLpt quark-gluon four vertex}

The quark-gluon four vertex with incoming quark momentum $r$, outgoing gluon momentum $p$, $q$, and outgoing quark momentum $s$ is
\bqa\label{4qg}
\Gamma_{abcd}^{\mu\nu,ij}(p,q,r,s)=-i g^2 \delta^{ij}m_q^2 \tilde{\mathcal{T}}^{\mu\nu}_{abcd}(p,q,r,s) \, ,
\eqa
and satisfies
\begin{equation}\label{4qg2}
\delta^{ij}\delta^{ad}\delta^{bc}\Gamma_{abcd,ij}^{\mu\nu}(p,q,r,s) =-4i g^2N_c d_A \Gamma^{\mu\nu}(p,q,r,s)\,,
\end{equation}
where
\bqa\label{4qg3}
\Gamma^{\mu\nu}(p,q,r,s) & = & m_q^2\bigg\langle y^\mu y^\nu  \bigg( \frac{1}{y\cdot r}+ \frac{1}{y\cdot s}   \bigg)  \nonumber \\ && \times \frac{{\displaystyle{\not}y}}{[y\cdot(r-p)][y\cdot(s+p)]}             \bigg\rangle \, .
\eqa
This tensor is symmetric in $\mu$ and $\nu$, and satisfies the Ward identity
\beq\label{4qg4}
p_\mu\Gamma^{\mu\nu}(p,q,r,s)= \Gamma^{\nu}(q,r-p,s)-\Gamma^{\nu}(q,r,s+p)\,.
\eeq

\subsection{HTLpt four scalar vertex}

The four-scalar vertex is
\bqa\label{4s}
&& \hspace{-8mm} \Gamma_{abcd}^{ABCD}(p,q,r,s)  =   -i g^2 \nonumber \\ && \times \bigg[ f_{abe}f_{cde}\bigg( \delta^{AC}\delta^{BD}-\delta^{AD}\delta^{BC}  \bigg) \nonumber \\ && +f_{ace}f_{bde}\bigg( \delta^{AB}\delta^{CD}-\delta^{AD}\delta^{BC}\bigg) \nonumber \\  && +f_{ade}f_{bce}\bigg( \delta^{AB}\delta^{CD}-\delta^{AC}\delta^{BD}    \bigg)   \bigg]\, .
\eqa
It satisfies
\beq\label{4s1}
\delta^{bd}\delta^{ac}\delta^{AC}\delta^{BD}\Gamma_{abcd}^{ABCD}(p,q,r,s)=(-i g^2)(60N_c d_A)\,.
\eeq

\subsection{HTLpt scalar-gluon three vertex}

The scalar-gluon vertex with incoming scalar momentum $r$, incoming gluon momentum $p$, and outgoing scalar momentum $q$ is
\beq\label{sg}
 \Gamma_{abc}^{\mu,AB}(p,q,r)=g f_{abc}\delta^{AB}(r+q)^\mu\,.
\eeq

\subsection{HTLpt scalar-gluon four vertex}

The scalar-gluon four vertex is independent of the direction of the momentum, and can be expressed as
\beq\label{ssgg}
 \Gamma_{abcde}^{\mu\nu,AB}(p,q,r,s)=-2ig^2 g^{\mu\nu} \delta^{AB}f_{ade}f_{bce} \, .
\eeq
It satisfies
\beq\label{ssgg1} \delta^{ac}\delta^{bd}\delta^{AB}\Gamma_{abcde}^{\mu\nu,AB}(p,q,r,s)=(2ig^2) (6N_c d_A)g^{\mu\nu}\,.
\eeq

\subsection{HTLpt quark-scalar vertex}

There are two types of quark-scalar vertices, the quark-scalar vertex and the quark-pseudoscalar vertex.  This is due to the fact that there are two types of interactions between quarks and scalars ($X_{\texttt{p}}$), and between quarks and pseudoscalars ($Y_{\texttt{q}}$) which are different. The quark-scalar vertex, with incoming scalar momentum $p$, incoming quark momentum $r$ and outgoing quark momentum $q$, and their corresponding colors indexed by $a,c,b$, can be written as
\beq\label{qs1}
 \Gamma_{abc,ij}^{\texttt{p}}(p,q,r)=-i g f_{abc}\alpha_{ij}^{\texttt{p}}\,.
\eeq
The quark-pseudoscalar vertex can be written as
\beq\label{qs1a}
 \Gamma_{abc,ij}^{\texttt{q}}(p,q,r)= g f_{abc}\beta_{ij}^{\texttt{q}}\gamma_5\,.
\eeq

\section{Integrals required in the HTLpt calculation}\label{IntegralHTLpt}

The one-loop sum-integrals required for the $\symff$ NLO HTLpt calculation were presented in Appendix B of Refs.~\cite{Andersen:2002ey,Andersen:2003zk}.  We list them here for completeness.
\begin{align}
&\sumint_{P} \frac{1}{P^2} = T^2 \bigg( \frac{\mu}{4\pi T} \bigg)^{2\epsilon}\frac{1}{12} \bigg[ 1+ \bigg( 2+ 2 \frac{\zeta'(-1)}{\zeta(-1)}\bigg)\epsilon  \bigg], \label{forlumab1} \\
&\sumint_{P} \frac{1}{P^4} = \frac{1}{(4\pi)^2} \bigg( \frac{\mu}{4\pi T} \bigg)^{2\epsilon} \bigg[ \frac{1}{\epsilon} +2 \gamma+ \bigg( \frac{\pi^2}{4}-4\gamma_1 \bigg)\epsilon  \bigg] , \label{forlumab2}\\
& \sumint_{P} \frac{p^2}{P^4}  = \frac{1}{8}T^2,\\
& \sumint_{P} \frac{p^2}{P^6}  = \frac{1}{(4\pi)^2}\bigg( \frac{\mu}{4\pi T} \bigg)^{2\epsilon} \frac{3}{4} \bigg[\frac{1}{\epsilon}+2\gamma -\frac{2}{3}  \bigg] ,\\
& \sumint_{\{P\}} \frac{p^2}{P^6}  = \frac{1}{(4\pi)^2}\bigg( \frac{\mu}{4\pi T} \bigg)^{2\epsilon} \frac{3}{4} \bigg[\frac{1}{\epsilon}+2\gamma -\frac{2}{3} +4 \log 2  \bigg] ,\\
& \sumint_{P} \frac{p^4}{P^8}  = \frac{1}{(4\pi)^2}\bigg( \frac{\mu}{4\pi T} \bigg)^{2\epsilon} \frac{5}{8} \bigg[\frac{1}{\epsilon}+2\gamma -\frac{16}{15}  \bigg] ,\\
& \sumint_{P} \frac{1}{P^2 p^2}  =   \frac{1}{(4\pi)^2}\bigg( \frac{\mu}{4\pi T} \bigg)^{2\epsilon} 2 \bigg[ \frac{1}{\epsilon} +2\gamma +2 \nonumber \\
& \hspace{2cm} + \bigg( 4 + 4 \gamma + \frac{\pi^2}{4} -4 \gamma_1 \bigg)\epsilon    \bigg] \, .
\end{align}
Similarly, the one-loop sum-integrals which involve HTL function ${\mathcal T}_P$ are given by
\begin{align}
& \sumint_{P} \frac{{\mathcal T}_P}{p^4} = \frac{1}{(4\pi)^2} \bigg( \frac{\mu}{4\pi T} \bigg)^{2\epsilon} (-1) \bigg[ \frac{1}{\epsilon} + 2 \gamma + 2 \log 2   \bigg], \\
& \sumint_{P} \frac{{\mathcal T}_P}{P^2 p^2} =   \frac{1}{(4\pi)^2}\bigg( \frac{\mu}{4\pi T} \bigg)^{2\epsilon} \bigg[ 2 \log 2 \bigg( \frac{1}{\epsilon} +2 \gamma \bigg) \nonumber \\
& \hspace{2cm} + 2 \log^2 2 +\frac{\pi^2}{3}     \bigg], \\
&\sumint_{P} \frac{({\mathcal T}_P)^2}{p^4} = \frac{1}{(4\pi)^2} \bigg( \frac{\mu}{4\pi T} \bigg)^{2\epsilon} \bigg( -\frac{2}{3} \bigg) \bigg[ (1+2 \log2) \nonumber \\
& \times \bigg( \frac{1}{\epsilon} + 2 \gamma  \bigg) - \frac{4}{3} + \frac{22}{3}\log2 + 2 \log^2 2 \bigg], \\
& \sumint_{\{P\}} \frac{{\mathcal T}_P}{P^2 P_0^2} =   \frac{1}{(4\pi)^2}\bigg( \frac{\mu}{4\pi T} \bigg)^{2\epsilon} \bigg[ \frac{1}{\epsilon^2} + 2(\gamma +2\log2 )\frac{1}{\epsilon} \nonumber \\
& \hspace{1cm} + \frac{\pi^2}{4} + 4\log^2 2 +8 \gamma \log2 -4\gamma_1 \bigg].
\end{align}
The two-loop sum-integrals required can be split into two types.  The first type are related to bosonic momentum integrations,
\begin{align}
& \sumint_{PQ} \frac{1}{P^2 Q^2 R^2}  =   0,  \\
& \sumint_{PQ} \frac{1}{P^2 Q^2 r^2}  =   \frac{T^2}{(4\pi)^2} \bigg( \frac{\mu}{4\pi T} \bigg)^{4\epsilon} \frac{1}{12} \bigg[ \frac{1}{\epsilon} +10 \nonumber \\
& \hspace{2cm} -12 \log2 + 4\frac{ \zeta'(-1)}{\zeta(-1)}   \bigg], \\
& \sumint_{PQ} \frac{q^2}{P^2 Q^2 r^4}  = \frac{T^2}{(4\pi)^2} \bigg( \frac{\mu}{4\pi T} \bigg)^{4\epsilon} \frac{1}{6} \bigg[ \frac{1}{\epsilon} + \frac{8}{3} \nonumber \\
& \hspace{2cm}  +2 \gamma +2 \frac{ \zeta'(-1)}{\zeta(-1)}  \bigg], \\
& \sumint_{PQ} \frac{q^2}{P^2 Q^2 R^2 r^2}  = \frac{T^2}{(4\pi)^2} \bigg( \frac{\mu}{4\pi T} \bigg)^{4\epsilon} \frac{1}{9} \bigg[ \frac{1}{\epsilon} + 7.521  \bigg], \\
& \sumint_{PQ} \frac{P\cdot Q}{P^2 Q^2 r^4}  = \frac{T^2}{(4\pi)^2} \bigg( \frac{\mu}{4\pi T} \bigg)^{4\epsilon} \bigg(-\frac{1}{8} \bigg) \bigg[ \frac{1}{\epsilon} + \frac{2}{9} \nonumber \\
& \hspace{2cm}  +4\log2 +\frac{8}{3}\gamma +\frac{4}{3}\frac{ \zeta'(-1)}{\zeta(-1)}   \bigg] .
\end{align}
The second type are related to fermionic momentum integrations,
\begin{align}
& \sumint_{\{PQ\}} \frac{1}{P^2 Q^2 R^2}  =   0,  \\
& \sumint_{\{PQ\}} \frac{1}{P^2 Q^2 r^2} = \frac{T^2}{(4\pi)^2} \bigg( \frac{\mu}{4\pi T} \bigg)^{4\epsilon} \bigg(-\frac{1}{6} \bigg) \bigg[ \frac{1}{\epsilon} \nonumber \\
& \hspace{1cm} + 4 -2\log 2 + 4\frac{ \zeta'(-1)}{\zeta(-1)}\bigg],
\end{align}
\begin{align}
& \sumint_{\{PQ\}} \frac{q^2}{P^2 Q^2 r^4} = \frac{T^2}{(4\pi)^2} \bigg( \frac{\mu}{4\pi T} \bigg)^{4\epsilon} \bigg(-\frac{1}{12} \bigg) \bigg[ \frac{1}{\epsilon} \nonumber \\
& \hspace{2cm} + \frac{11}{3} + 2\gamma -2\log 2 + 2\frac{ \zeta'(-1)}{\zeta(-1)}\bigg],\\
& \sumint_{\{PQ\}} \frac{P\cdot Q}{P^2 Q^2 r^4} = \frac{T^2}{(4\pi)^2} \bigg( \frac{\mu}{4\pi T} \bigg)^{4\epsilon} \bigg(-\frac{1}{36} \bigg) \nonumber \\
& \hspace{2cm} \times \bigg[ 1-6\gamma + 6\frac{ \zeta'(-1)}{\zeta(-1)}\bigg],\\
& \sumint_{\{PQ\}} \frac{q^2}{P^2 Q^2 R^2 r^2}  = \frac{T^2}{(4\pi)^2} \bigg( \frac{\mu}{4\pi T} \bigg)^{4\epsilon} \bigg(-\frac{1}{72} \bigg) \nonumber \\
& \hspace{3cm} \times \bigg[ \frac{1}{\epsilon} - 7.002  \bigg], \\
& \sumint_{\{PQ\}} \frac{p^2}{P^2 Q^2 R^2 q^2}  = \frac{T^2}{(4\pi)^2} \bigg( \frac{\mu}{4\pi T} \bigg)^{4\epsilon} \frac{5}{72} \bigg[ \frac{1}{\epsilon} + 9.5667  \bigg], \\
& \sumint_{\{PQ\}} \frac{r^2}{P^2 Q^2 R^2 q^2}  = \frac{T^2}{(4\pi)^2} \bigg( \frac{\mu}{4\pi T} \bigg)^{4\epsilon} \bigg(-\frac{1}{18} \bigg) \nonumber \\
& \hspace{3cm}\times \bigg[ \frac{1}{\epsilon} + 8.1420  \bigg].
\end{align}
In a similar manner, the two-loop sum-integrals involving ${\mathcal T}_P$ can be split into two types.  The first type are related to the bosonic momentum integrations,
\begin{align}
& \sumint_{PQ} \frac{{\mathcal T}_R}{P^2 Q^2 r^2}  =   \frac{T^2}{(4\pi)^2}\bigg( \frac{\mu}{4\pi T} \bigg)^{4\epsilon}  \bigg(-\frac{1}{48} \bigg) \bigg[ \frac{1}{\epsilon^2} \nonumber \\
& \hspace{0.5cm} + \bigg( 2-12 \log2 +4 \frac{ \zeta'(-1)}{\zeta(-1)} \bigg)\frac{1}{\epsilon} -19.83   \bigg],\\
& \sumint_{PQ} \frac{q^2 {\mathcal T}_R}{P^2 Q^2 r^4}  =   \frac{T^2}{(4\pi)^2}\bigg( \frac{\mu}{4\pi T} \bigg)^{4\epsilon}  \bigg(-\frac{1}{576} \bigg) \bigg[ \frac{1}{\epsilon^2} \nonumber \\
& + \bigg( \frac{26}{3} - \frac{24}{\pi^2} -92\log2   +4\frac{ \zeta'(-1)}{\zeta(-1)}   \bigg)\frac{1}{\epsilon} - 477.7   \bigg],\\
& \sumint_{PQ} \frac{(P\cdot Q) {\mathcal T}_R}{P^2 Q^2 r^4}  =   \frac{T^2}{(4\pi)^2}\bigg( \frac{\mu}{4\pi T} \bigg)^{4\epsilon}  \bigg(-\frac{1}{96} \bigg) \bigg[ \frac{1}{\epsilon^2} \nonumber \\
& \hspace{0.5cm} + \bigg( \frac{8}{\pi^2} +4\log2 +4\frac{ \zeta'(-1)}{\zeta(-1)}  \bigg)\frac{1}{\epsilon} + 59.66   \bigg].
\end{align}
The second type are related to the fermionic momentum integrations,
\begin{align}
& \sumint_{\{PQ\}} \frac{{\mathcal T}_R}{P^2 Q^2 r^2} = \frac{T^2}{(4\pi)^2} \bigg( \frac{\mu}{4\pi T} \bigg)^{4\epsilon} \bigg(-\frac{1}{48} \bigg)  \bigg[  \frac{1}{\epsilon^2} \nonumber \\
& \hspace{0.5cm} + \bigg( 2  + 12\log2   + 4 \frac{ \zeta'(-1)}{\zeta(-1)}    \bigg) \frac{1}{\epsilon}  + 136.362 \bigg],\\
& \sumint_{\{PQ\}} \frac{q^2{\mathcal T}_R}{P^2 Q^2 r^4} = \frac{T^2}{(4\pi)^2} \bigg( \frac{\mu}{4\pi T} \bigg)^{4\epsilon} \bigg(-\frac{1}{576} \bigg)  \bigg[  \frac{1}{\epsilon^2}  \nonumber \\ &\hspace{0.5cm} + \bigg( \frac{26}{3} + 52\log2 + 4 \frac{ \zeta'(-1)}{\zeta(-1)}    \bigg) \frac{1}{\epsilon}  + 446.438 \bigg],
\end{align}

\begin{align}
& \sumint_{\{PQ\}} \frac{(P\cdot Q){\mathcal T}_R}{P^2 Q^2 r^4} = \frac{T^2}{(4\pi)^2} \bigg( \frac{\mu}{4\pi T} \bigg)^{4\epsilon} \bigg(-\frac{1}{96} \bigg)  \bigg[  \frac{1}{\epsilon^2} 
\nonumber \\ &  \hspace{0.5cm} 
+ \bigg( 4\log2   + 4 \frac{ \zeta'(-1)}{\zeta(-1)}    \bigg) \frac{1}{\epsilon}  + 69.174 \bigg],\\
& \sumint_{\{PQ\}} \frac{(r^2-p^2){\mathcal T}_Q}{P^2 q^2 Q_0^2 R^2}  =  - \frac{T^2}{(4\pi)^2}\bigg( \frac{\mu}{4\pi T} \bigg)^{4\epsilon}  \frac{1}{8} \bigg[ \frac{1}{\epsilon^2} \nonumber \\
&  + \bigg( 2+ 2 \gamma +\frac{10}{3}\log2  + 2 \frac{ \zeta'(-1)}{\zeta(-1)} \bigg)\frac{1}{\epsilon} +46.8757   \bigg].
\end{align}
%

\bibliographystyle{apsrev4-1}
\bibliography{scheme}

\begin{thebibliography}{71}%
\makeatletter
\providecommand \@ifxundefined [1]{%
 \@ifx{#1\undefined}
}%
\providecommand \@ifnum [1]{%
 \ifnum #1\expandafter \@firstoftwo
 \else \expandafter \@secondoftwo
 \fi
}%
\providecommand \@ifx [1]{%
 \ifx #1\expandafter \@firstoftwo
 \else \expandafter \@secondoftwo
 \fi
}%
\providecommand \natexlab [1]{#1}%
\providecommand \enquote  [1]{``#1''}%
\providecommand \bibnamefont  [1]{#1}%
\providecommand \bibfnamefont [1]{#1}%
\providecommand \citenamefont [1]{#1}%
\providecommand \href@noop [0]{\@secondoftwo}%
\providecommand \href [0]{\begingroup \@sanitize@url \@href}%
\providecommand \@href[1]{\@@startlink{#1}\@@href}%
\providecommand \@@href[1]{\endgroup#1\@@endlink}%
\providecommand \@sanitize@url [0]{\catcode `\\12\catcode `\$12\catcode
  `\&12\catcode `\#12\catcode `\^12\catcode `\_12\catcode `\%12\relax}%
\providecommand \@@startlink[1]{}%
\providecommand \@@endlink[0]{}%
\providecommand \url  [0]{\begingroup\@sanitize@url \@url }%
\providecommand \@url [1]{\endgroup\@href {#1}{\urlprefix }}%
\providecommand \urlprefix  [0]{URL }%
\providecommand \Eprint [0]{\href }%
\providecommand \doibase [0]{http://dx.doi.org/}%
\providecommand \selectlanguage [0]{\@gobble}%
\providecommand \bibinfo  [0]{\@secondoftwo}%
\providecommand \bibfield  [0]{\@secondoftwo}%
\providecommand \translation [1]{[#1]}%
\providecommand \BibitemOpen [0]{}%
\providecommand \bibitemStop [0]{}%
\providecommand \bibitemNoStop [0]{.\EOS\space}%
\providecommand \EOS [0]{\spacefactor3000\relax}%
\providecommand \BibitemShut  [1]{\csname bibitem#1\endcsname}%
\let\auto@bib@innerbib\@empty
\bibitem [{\citenamefont {Gervais}\ and\ \citenamefont
  {Sakita}(1971)}]{Gervais:1971ji}%
  \BibitemOpen
  \bibfield  {author} {\bibinfo {author} {\bibfnamefont {J.-L.}\ \bibnamefont
  {Gervais}}\ and\ \bibinfo {author} {\bibfnamefont {B.}~\bibnamefont
  {Sakita}},\ }\href {\doibase 10.1016/0550-3213(71)90351-8} {\bibfield
  {journal} {\bibinfo  {journal} {Nucl. Phys. B}\ }\textbf {\bibinfo {volume}
  {34}},\ \bibinfo {pages} {632} (\bibinfo {year} {1971})}\BibitemShut
  {NoStop}%
\bibitem [{\citenamefont {Volkov}\ and\ \citenamefont
  {Akulov}(1973)}]{Volkov:1973ix}%
  \BibitemOpen
  \bibfield  {author} {\bibinfo {author} {\bibfnamefont {D.~V.}\ \bibnamefont
  {Volkov}}\ and\ \bibinfo {author} {\bibfnamefont {V.~P.}\ \bibnamefont
  {Akulov}},\ }\href {\doibase 10.1016/0370-2693(73)90490-5} {\bibfield
  {journal} {\bibinfo  {journal} {Phys. Lett. B}\ }\textbf {\bibinfo {volume}
  {46}},\ \bibinfo {pages} {109} (\bibinfo {year} {1973})}\BibitemShut
  {NoStop}%
\bibitem [{\citenamefont {Akulov}\ and\ \citenamefont
  {Volkov}(1974)}]{Akulov:1974xz}%
  \BibitemOpen
  \bibfield  {author} {\bibinfo {author} {\bibfnamefont {V.~P.}\ \bibnamefont
  {Akulov}}\ and\ \bibinfo {author} {\bibfnamefont {D.~V.}\ \bibnamefont
  {Volkov}},\ }\href {\doibase 10.1007/BF01036922} {\bibfield  {journal}
  {\bibinfo  {journal} {Theor. Math. Phys.}\ }\textbf {\bibinfo {volume}
  {18}},\ \bibinfo {pages} {28} (\bibinfo {year} {1974})}\BibitemShut {NoStop}%
\bibitem [{\citenamefont {Nilles}(1984)}]{Nilles:1983ge}%
  \BibitemOpen
  \bibfield  {author} {\bibinfo {author} {\bibfnamefont {H.~P.}\ \bibnamefont
  {Nilles}},\ }\href {\doibase 10.1016/0370-1573(84)90008-5} {\bibfield
  {journal} {\bibinfo  {journal} {Phys. Rept.}\ }\textbf {\bibinfo {volume}
  {110}},\ \bibinfo {pages} {1} (\bibinfo {year} {1984})}\BibitemShut {NoStop}%
\bibitem [{\citenamefont {Martin}(1998)}]{Martin:1997ns}%
  \BibitemOpen
  \bibfield  {author} {\bibinfo {author} {\bibfnamefont {S.~P.}\ \bibnamefont
  {Martin}},\ }\href {\doibase 10.1142/9789812839657_0001} {\bibfield
  {journal} {\bibinfo  {journal} {Adv. Ser. Direct. High Energy Phys.}\
  }\textbf {\bibinfo {volume} {18}},\ \bibinfo {pages} {1} (\bibinfo {year}
  {1998})},\ \Eprint {http://arxiv.org/abs/hep-ph/9709356}
  {arXiv:hep-ph/9709356} \BibitemShut {NoStop}%
\bibitem [{\citenamefont {Maldacena}(1998)}]{Maldacena:1997re}%
  \BibitemOpen
  \bibfield  {author} {\bibinfo {author} {\bibfnamefont {J.~M.}\ \bibnamefont
  {Maldacena}},\ }\href {\doibase 10.1023/A:1026654312961} {\bibfield
  {journal} {\bibinfo  {journal} {Adv. Theor. Math. Phys.}\ }\textbf {\bibinfo
  {volume} {2}},\ \bibinfo {pages} {231} (\bibinfo {year} {1998})},\ \Eprint
  {http://arxiv.org/abs/hep-th/9711200} {arXiv:hep-th/9711200} \BibitemShut
  {NoStop}%
\bibitem [{\citenamefont {Gubser}\ \emph {et~al.}(1998)\citenamefont {Gubser},
  \citenamefont {Klebanov},\ and\ \citenamefont {Tseytlin}}]{Gubser:1998nz}%
  \BibitemOpen
  \bibfield  {author} {\bibinfo {author} {\bibfnamefont {S.~S.}\ \bibnamefont
  {Gubser}}, \bibinfo {author} {\bibfnamefont {I.~R.}\ \bibnamefont
  {Klebanov}}, \ and\ \bibinfo {author} {\bibfnamefont {A.~A.}\ \bibnamefont
  {Tseytlin}},\ }\href {\doibase 10.1016/S0550-3213(98)00514-8} {\bibfield
  {journal} {\bibinfo  {journal} {Nucl. Phys.}\ }\textbf {\bibinfo {volume}
  {B534}},\ \bibinfo {pages} {202} (\bibinfo {year} {1998})},\ \Eprint
  {http://arxiv.org/abs/hep-th/9805156} {arXiv:hep-th/9805156 [hep-th]}
  \BibitemShut {NoStop}%
\bibitem [{\citenamefont {Andersen}\ \emph {et~al.}(1999)\citenamefont
  {Andersen}, \citenamefont {Braaten},\ and\ \citenamefont
  {Strickland}}]{Andersen:1999fw}%
  \BibitemOpen
  \bibfield  {author} {\bibinfo {author} {\bibfnamefont {J.~O.}\ \bibnamefont
  {Andersen}}, \bibinfo {author} {\bibfnamefont {E.}~\bibnamefont {Braaten}}, \
  and\ \bibinfo {author} {\bibfnamefont {M.}~\bibnamefont {Strickland}},\
  }\href {\doibase 10.1103/PhysRevLett.83.2139} {\bibfield  {journal} {\bibinfo
   {journal} {Phys. Rev. Lett.}\ }\textbf {\bibinfo {volume} {83}},\ \bibinfo
  {pages} {2139} (\bibinfo {year} {1999})},\ \Eprint
  {http://arxiv.org/abs/hep-ph/9902327} {arXiv:hep-ph/9902327 [hep-ph]}
  \BibitemShut {NoStop}%
\bibitem [{\citenamefont {Blaizot}\ \emph
  {et~al.}(1999{\natexlab{a}})\citenamefont {Blaizot}, \citenamefont {Iancu},\
  and\ \citenamefont {Rebhan}}]{Blaizot:1999ip}%
  \BibitemOpen
  \bibfield  {author} {\bibinfo {author} {\bibfnamefont {J.~P.}\ \bibnamefont
  {Blaizot}}, \bibinfo {author} {\bibfnamefont {E.}~\bibnamefont {Iancu}}, \
  and\ \bibinfo {author} {\bibfnamefont {A.}~\bibnamefont {Rebhan}},\ }\href
  {\doibase 10.1103/PhysRevLett.83.2906} {\bibfield  {journal} {\bibinfo
  {journal} {Phys. Rev. Lett.}\ }\textbf {\bibinfo {volume} {83}},\ \bibinfo
  {pages} {2906} (\bibinfo {year} {1999}{\natexlab{a}})},\ \Eprint
  {http://arxiv.org/abs/hep-ph/9906340} {arXiv:hep-ph/9906340 [hep-ph]}
  \BibitemShut {NoStop}%
\bibitem [{\citenamefont {Andersen}\ \emph
  {et~al.}(2000{\natexlab{a}})\citenamefont {Andersen}, \citenamefont
  {Braaten},\ and\ \citenamefont {Strickland}}]{Andersen:1999sf}%
  \BibitemOpen
  \bibfield  {author} {\bibinfo {author} {\bibfnamefont {J.~O.}\ \bibnamefont
  {Andersen}}, \bibinfo {author} {\bibfnamefont {E.}~\bibnamefont {Braaten}}, \
  and\ \bibinfo {author} {\bibfnamefont {M.}~\bibnamefont {Strickland}},\
  }\href {\doibase 10.1103/PhysRevD.61.014017} {\bibfield  {journal} {\bibinfo
  {journal} {Phys. Rev.}\ }\textbf {\bibinfo {volume} {D61}},\ \bibinfo {pages}
  {014017} (\bibinfo {year} {2000}{\natexlab{a}})},\ \Eprint
  {http://arxiv.org/abs/hep-ph/9905337} {arXiv:hep-ph/9905337 [hep-ph]}
  \BibitemShut {NoStop}%
\bibitem [{\citenamefont {Blaizot}\ \emph
  {et~al.}(1999{\natexlab{b}})\citenamefont {Blaizot}, \citenamefont {Iancu},\
  and\ \citenamefont {Rebhan}}]{Blaizot:1999ap}%
  \BibitemOpen
  \bibfield  {author} {\bibinfo {author} {\bibfnamefont {J.~P.}\ \bibnamefont
  {Blaizot}}, \bibinfo {author} {\bibfnamefont {E.}~\bibnamefont {Iancu}}, \
  and\ \bibinfo {author} {\bibfnamefont {A.}~\bibnamefont {Rebhan}},\ }\href
  {\doibase 10.1016/S0370-2693(99)01306-4} {\bibfield  {journal} {\bibinfo
  {journal} {Phys. Lett.}\ }\textbf {\bibinfo {volume} {B470}},\ \bibinfo
  {pages} {181} (\bibinfo {year} {1999}{\natexlab{b}})},\ \Eprint
  {http://arxiv.org/abs/hep-ph/9910309} {arXiv:hep-ph/9910309 [hep-ph]}
  \BibitemShut {NoStop}%
\bibitem [{\citenamefont {Andersen}\ \emph
  {et~al.}(2000{\natexlab{b}})\citenamefont {Andersen}, \citenamefont
  {Braaten},\ and\ \citenamefont {Strickland}}]{Andersen:1999va}%
  \BibitemOpen
  \bibfield  {author} {\bibinfo {author} {\bibfnamefont {J.~O.}\ \bibnamefont
  {Andersen}}, \bibinfo {author} {\bibfnamefont {E.}~\bibnamefont {Braaten}}, \
  and\ \bibinfo {author} {\bibfnamefont {M.}~\bibnamefont {Strickland}},\
  }\href {\doibase 10.1103/PhysRevD.61.074016} {\bibfield  {journal} {\bibinfo
  {journal} {Phys. Rev.}\ }\textbf {\bibinfo {volume} {D61}},\ \bibinfo {pages}
  {074016} (\bibinfo {year} {2000}{\natexlab{b}})},\ \Eprint
  {http://arxiv.org/abs/hep-ph/9908323} {arXiv:hep-ph/9908323 [hep-ph]}
  \BibitemShut {NoStop}%
\bibitem [{\citenamefont {Blaizot}\ \emph
  {et~al.}(2001{\natexlab{a}})\citenamefont {Blaizot}, \citenamefont {Iancu},\
  and\ \citenamefont {Rebhan}}]{Blaizot:2000fc}%
  \BibitemOpen
  \bibfield  {author} {\bibinfo {author} {\bibfnamefont {J.~P.}\ \bibnamefont
  {Blaizot}}, \bibinfo {author} {\bibfnamefont {E.}~\bibnamefont {Iancu}}, \
  and\ \bibinfo {author} {\bibfnamefont {A.}~\bibnamefont {Rebhan}},\ }\href
  {\doibase 10.1103/PhysRevD.63.065003} {\bibfield  {journal} {\bibinfo
  {journal} {Phys. Rev.}\ }\textbf {\bibinfo {volume} {D63}},\ \bibinfo {pages}
  {065003} (\bibinfo {year} {2001}{\natexlab{a}})},\ \Eprint
  {http://arxiv.org/abs/hep-ph/0005003} {arXiv:hep-ph/0005003 [hep-ph]}
  \BibitemShut {NoStop}%
\bibitem [{\citenamefont {Peshier}(2001)}]{Peshier:2000hx}%
  \BibitemOpen
  \bibfield  {author} {\bibinfo {author} {\bibfnamefont {A.}~\bibnamefont
  {Peshier}},\ }\href {\doibase 10.1103/PhysRevD.63.105004} {\bibfield
  {journal} {\bibinfo  {journal} {Phys. Rev.}\ }\textbf {\bibinfo {volume}
  {D63}},\ \bibinfo {pages} {105004} (\bibinfo {year} {2001})},\ \Eprint
  {http://arxiv.org/abs/hep-ph/0011250} {arXiv:hep-ph/0011250 [hep-ph]}
  \BibitemShut {NoStop}%
\bibitem [{\citenamefont {Blaizot}\ \emph
  {et~al.}(2001{\natexlab{b}})\citenamefont {Blaizot}, \citenamefont {Iancu},\
  and\ \citenamefont {Rebhan}}]{Blaizot:2001vr}%
  \BibitemOpen
  \bibfield  {author} {\bibinfo {author} {\bibfnamefont {J.~P.}\ \bibnamefont
  {Blaizot}}, \bibinfo {author} {\bibfnamefont {E.}~\bibnamefont {Iancu}}, \
  and\ \bibinfo {author} {\bibfnamefont {A.}~\bibnamefont {Rebhan}},\ }\href
  {\doibase 10.1016/S0370-2693(01)01316-8} {\bibfield  {journal} {\bibinfo
  {journal} {Phys. Lett. B}\ }\textbf {\bibinfo {volume} {523}},\ \bibinfo
  {pages} {143} (\bibinfo {year} {2001}{\natexlab{b}})},\ \Eprint
  {http://arxiv.org/abs/hep-ph/0110369} {arXiv:hep-ph/0110369} \BibitemShut
  {NoStop}%
\bibitem [{\citenamefont {Andersen}\ \emph {et~al.}(2002)\citenamefont
  {Andersen}, \citenamefont {Braaten}, \citenamefont {Petitgirard},\ and\
  \citenamefont {Strickland}}]{Andersen:2002ey}%
  \BibitemOpen
  \bibfield  {author} {\bibinfo {author} {\bibfnamefont {J.~O.}\ \bibnamefont
  {Andersen}}, \bibinfo {author} {\bibfnamefont {E.}~\bibnamefont {Braaten}},
  \bibinfo {author} {\bibfnamefont {E.}~\bibnamefont {Petitgirard}}, \ and\
  \bibinfo {author} {\bibfnamefont {M.}~\bibnamefont {Strickland}},\ }\href
  {\doibase 10.1103/PhysRevD.66.085016} {\bibfield  {journal} {\bibinfo
  {journal} {Phys. Rev.}\ }\textbf {\bibinfo {volume} {D66}},\ \bibinfo {pages}
  {085016} (\bibinfo {year} {2002})},\ \Eprint
  {http://arxiv.org/abs/hep-ph/0205085} {arXiv:hep-ph/0205085 [hep-ph]}
  \BibitemShut {NoStop}%
\bibitem [{\citenamefont {Andersen}\ \emph {et~al.}(2004)\citenamefont
  {Andersen}, \citenamefont {Petitgirard},\ and\ \citenamefont
  {Strickland}}]{Andersen:2003zk}%
  \BibitemOpen
  \bibfield  {author} {\bibinfo {author} {\bibfnamefont {J.~O.}\ \bibnamefont
  {Andersen}}, \bibinfo {author} {\bibfnamefont {E.}~\bibnamefont
  {Petitgirard}}, \ and\ \bibinfo {author} {\bibfnamefont {M.}~\bibnamefont
  {Strickland}},\ }\href {\doibase 10.1103/PhysRevD.70.045001} {\bibfield
  {journal} {\bibinfo  {journal} {Phys. Rev.}\ }\textbf {\bibinfo {volume}
  {D70}},\ \bibinfo {pages} {045001} (\bibinfo {year} {2004})},\ \Eprint
  {http://arxiv.org/abs/hep-ph/0302069} {arXiv:hep-ph/0302069 [hep-ph]}
  \BibitemShut {NoStop}%
\bibitem [{\citenamefont {Blaizot}\ \emph {et~al.}(2003)\citenamefont
  {Blaizot}, \citenamefont {Iancu},\ and\ \citenamefont
  {Rebhan}}]{Blaizot:2003iq}%
  \BibitemOpen
  \bibfield  {author} {\bibinfo {author} {\bibfnamefont {J.~P.}\ \bibnamefont
  {Blaizot}}, \bibinfo {author} {\bibfnamefont {E.}~\bibnamefont {Iancu}}, \
  and\ \bibinfo {author} {\bibfnamefont {A.}~\bibnamefont {Rebhan}},\ }\href
  {\doibase 10.1103/PhysRevD.68.025011} {\bibfield  {journal} {\bibinfo
  {journal} {Phys. Rev. D}\ }\textbf {\bibinfo {volume} {68}},\ \bibinfo
  {pages} {025011} (\bibinfo {year} {2003})},\ \Eprint
  {http://arxiv.org/abs/hep-ph/0303045} {arXiv:hep-ph/0303045} \BibitemShut
  {NoStop}%
\bibitem [{\citenamefont {Andersen}\ \emph
  {et~al.}(2010{\natexlab{a}})\citenamefont {Andersen}, \citenamefont
  {Strickland},\ and\ \citenamefont {Su}}]{Andersen:2009tc}%
  \BibitemOpen
  \bibfield  {author} {\bibinfo {author} {\bibfnamefont {J.~O.}\ \bibnamefont
  {Andersen}}, \bibinfo {author} {\bibfnamefont {M.}~\bibnamefont
  {Strickland}}, \ and\ \bibinfo {author} {\bibfnamefont {N.}~\bibnamefont
  {Su}},\ }\href {\doibase 10.1103/PhysRevLett.104.122003} {\bibfield
  {journal} {\bibinfo  {journal} {Phys.\ Rev.\ Lett.}\ }\textbf {\bibinfo
  {volume} {104}},\ \bibinfo {pages} {122003} (\bibinfo {year}
  {2010}{\natexlab{a}})},\ \Eprint {http://arxiv.org/abs/0911.0676}
  {arXiv:0911.0676 [hep-ph]} \BibitemShut {NoStop}%
\bibitem [{\citenamefont {Andersen}\ \emph
  {et~al.}(2010{\natexlab{b}})\citenamefont {Andersen}, \citenamefont
  {Strickland},\ and\ \citenamefont {Su}}]{Andersen:2010ct}%
  \BibitemOpen
  \bibfield  {author} {\bibinfo {author} {\bibfnamefont {J.~O.}\ \bibnamefont
  {Andersen}}, \bibinfo {author} {\bibfnamefont {M.}~\bibnamefont
  {Strickland}}, \ and\ \bibinfo {author} {\bibfnamefont {N.}~\bibnamefont
  {Su}},\ }\href {\doibase 10.1007/JHEP08(2010)113} {\bibfield  {journal}
  {\bibinfo  {journal} {JHEP}\ }\textbf {\bibinfo {volume} {08}},\ \bibinfo
  {pages} {113} (\bibinfo {year} {2010}{\natexlab{b}})},\ \Eprint
  {http://arxiv.org/abs/1005.1603} {arXiv:1005.1603 [hep-ph]} \BibitemShut
  {NoStop}%
\bibitem [{\citenamefont {Andersen}\ \emph
  {et~al.}(2011{\natexlab{a}})\citenamefont {Andersen}, \citenamefont
  {Leganger}, \citenamefont {Strickland},\ and\ \citenamefont
  {Su}}]{Andersen:2010wu}%
  \BibitemOpen
  \bibfield  {author} {\bibinfo {author} {\bibfnamefont {J.~O.}\ \bibnamefont
  {Andersen}}, \bibinfo {author} {\bibfnamefont {L.~E.}\ \bibnamefont
  {Leganger}}, \bibinfo {author} {\bibfnamefont {M.}~\bibnamefont
  {Strickland}}, \ and\ \bibinfo {author} {\bibfnamefont {N.}~\bibnamefont
  {Su}},\ }\href {\doibase 10.1016/j.physletb.2010.12.070} {\bibfield
  {journal} {\bibinfo  {journal} {Phys.\ Lett.\ B}\ }\textbf {\bibinfo {volume}
  {696}},\ \bibinfo {pages} {468} (\bibinfo {year} {2011}{\natexlab{a}})},\
  \Eprint {http://arxiv.org/abs/1009.4644} {arXiv:1009.4644 [hep-ph]}
  \BibitemShut {NoStop}%
\bibitem [{\citenamefont {Andersen}\ \emph
  {et~al.}(2011{\natexlab{b}})\citenamefont {Andersen}, \citenamefont
  {Leganger}, \citenamefont {Strickland},\ and\ \citenamefont
  {Su}}]{Andersen:2011sf}%
  \BibitemOpen
  \bibfield  {author} {\bibinfo {author} {\bibfnamefont {J.~O.}\ \bibnamefont
  {Andersen}}, \bibinfo {author} {\bibfnamefont {L.~E.}\ \bibnamefont
  {Leganger}}, \bibinfo {author} {\bibfnamefont {M.}~\bibnamefont
  {Strickland}}, \ and\ \bibinfo {author} {\bibfnamefont {N.}~\bibnamefont
  {Su}},\ }\href {\doibase 10.1007/JHEP08(2011)053} {\bibfield  {journal}
  {\bibinfo  {journal} {JHEP}\ }\textbf {\bibinfo {volume} {08}},\ \bibinfo
  {pages} {053} (\bibinfo {year} {2011}{\natexlab{b}})},\ \Eprint
  {http://arxiv.org/abs/1103.2528} {arXiv:1103.2528 [hep-ph]} \BibitemShut
  {NoStop}%
\bibitem [{\citenamefont {Andersen}\ \emph
  {et~al.}(2011{\natexlab{c}})\citenamefont {Andersen}, \citenamefont
  {Leganger}, \citenamefont {Strickland},\ and\ \citenamefont
  {Su}}]{Andersen:2011ug}%
  \BibitemOpen
  \bibfield  {author} {\bibinfo {author} {\bibfnamefont {J.~O.}\ \bibnamefont
  {Andersen}}, \bibinfo {author} {\bibfnamefont {L.~E.}\ \bibnamefont
  {Leganger}}, \bibinfo {author} {\bibfnamefont {M.}~\bibnamefont
  {Strickland}}, \ and\ \bibinfo {author} {\bibfnamefont {N.}~\bibnamefont
  {Su}},\ }\href {\doibase 10.1103/PhysRevD.84.087703} {\bibfield  {journal}
  {\bibinfo  {journal} {Phys.\ Rev.\ D}\ }\textbf {\bibinfo {volume} {84}},\
  \bibinfo {pages} {087703} (\bibinfo {year} {2011}{\natexlab{c}})},\ \Eprint
  {http://arxiv.org/abs/1106.0514} {arXiv:1106.0514 [hep-ph]} \BibitemShut
  {NoStop}%
\bibitem [{\citenamefont {Haque}\ \emph
  {et~al.}(2013{\natexlab{a}})\citenamefont {Haque}, \citenamefont {Mustafa},\
  and\ \citenamefont {Strickland}}]{Haque:2012my}%
  \BibitemOpen
  \bibfield  {author} {\bibinfo {author} {\bibfnamefont {N.}~\bibnamefont
  {Haque}}, \bibinfo {author} {\bibfnamefont {M.~G.}\ \bibnamefont {Mustafa}},
  \ and\ \bibinfo {author} {\bibfnamefont {M.}~\bibnamefont {Strickland}},\
  }\href {\doibase 10.1103/PhysRevD.87.105007} {\bibfield  {journal} {\bibinfo
  {journal} {Phys. Rev. D}\ }\textbf {\bibinfo {volume} {87}},\ \bibinfo
  {pages} {105007} (\bibinfo {year} {2013}{\natexlab{a}})},\ \Eprint
  {http://arxiv.org/abs/1212.1797} {arXiv:1212.1797 [hep-ph]} \BibitemShut
  {NoStop}%
\bibitem [{\citenamefont {Haque}\ \emph
  {et~al.}(2013{\natexlab{b}})\citenamefont {Haque}, \citenamefont {Mustafa},\
  and\ \citenamefont {Strickland}}]{Haque:2013qta}%
  \BibitemOpen
  \bibfield  {author} {\bibinfo {author} {\bibfnamefont {N.}~\bibnamefont
  {Haque}}, \bibinfo {author} {\bibfnamefont {M.~G.}\ \bibnamefont {Mustafa}},
  \ and\ \bibinfo {author} {\bibfnamefont {M.}~\bibnamefont {Strickland}},\
  }\href {\doibase 10.1007/JHEP07(2013)184} {\bibfield  {journal} {\bibinfo
  {journal} {JHEP}\ }\textbf {\bibinfo {volume} {07}},\ \bibinfo {pages} {184}
  (\bibinfo {year} {2013}{\natexlab{b}})},\ \Eprint
  {http://arxiv.org/abs/1302.3228} {arXiv:1302.3228 [hep-ph]} \BibitemShut
  {NoStop}%
\bibitem [{\citenamefont {Haque}\ \emph
  {et~al.}(2014{\natexlab{a}})\citenamefont {Haque}, \citenamefont {Andersen},
  \citenamefont {Mustafa}, \citenamefont {Strickland},\ and\ \citenamefont
  {Su}}]{Haque:2013sja}%
  \BibitemOpen
  \bibfield  {author} {\bibinfo {author} {\bibfnamefont {N.}~\bibnamefont
  {Haque}}, \bibinfo {author} {\bibfnamefont {J.~O.}\ \bibnamefont {Andersen}},
  \bibinfo {author} {\bibfnamefont {M.~G.}\ \bibnamefont {Mustafa}}, \bibinfo
  {author} {\bibfnamefont {M.}~\bibnamefont {Strickland}}, \ and\ \bibinfo
  {author} {\bibfnamefont {N.}~\bibnamefont {Su}},\ }\href {\doibase
  10.1103/PhysRevD.89.061701} {\bibfield  {journal} {\bibinfo  {journal}
  {Phys.\ Rev.\ D}\ }\textbf {\bibinfo {volume} {89}},\ \bibinfo {pages}
  {061701} (\bibinfo {year} {2014}{\natexlab{a}})},\ \Eprint
  {http://arxiv.org/abs/1309.3968} {arXiv:1309.3968 [hep-ph]} \BibitemShut
  {NoStop}%
\bibitem [{\citenamefont {Haque}\ \emph
  {et~al.}(2014{\natexlab{b}})\citenamefont {Haque}, \citenamefont
  {Bandyopadhyay}, \citenamefont {Andersen}, \citenamefont {Mustafa},
  \citenamefont {Strickland},\ and\ \citenamefont {Su}}]{Haque:2014rua}%
  \BibitemOpen
  \bibfield  {author} {\bibinfo {author} {\bibfnamefont {N.}~\bibnamefont
  {Haque}}, \bibinfo {author} {\bibfnamefont {A.}~\bibnamefont
  {Bandyopadhyay}}, \bibinfo {author} {\bibfnamefont {J.~O.}\ \bibnamefont
  {Andersen}}, \bibinfo {author} {\bibfnamefont {M.~G.}\ \bibnamefont
  {Mustafa}}, \bibinfo {author} {\bibfnamefont {M.}~\bibnamefont {Strickland}},
  \ and\ \bibinfo {author} {\bibfnamefont {N.}~\bibnamefont {Su}},\ }\href
  {\doibase 10.1007/JHEP05(2014)027} {\bibfield  {journal} {\bibinfo  {journal}
  {JHEP}\ }\textbf {\bibinfo {volume} {05}},\ \bibinfo {pages} {027} (\bibinfo
  {year} {2014}{\natexlab{b}})},\ \Eprint {http://arxiv.org/abs/1402.6907}
  {arXiv:1402.6907 [hep-ph]} \BibitemShut {NoStop}%
\bibitem [{\citenamefont {Andersen}\ \emph {et~al.}(2016)\citenamefont
  {Andersen}, \citenamefont {Haque}, \citenamefont {Mustafa},\ and\
  \citenamefont {Strickland}}]{Andersen:2015eoa}%
  \BibitemOpen
  \bibfield  {author} {\bibinfo {author} {\bibfnamefont {J.~O.}\ \bibnamefont
  {Andersen}}, \bibinfo {author} {\bibfnamefont {N.}~\bibnamefont {Haque}},
  \bibinfo {author} {\bibfnamefont {M.~G.}\ \bibnamefont {Mustafa}}, \ and\
  \bibinfo {author} {\bibfnamefont {M.}~\bibnamefont {Strickland}},\ }\href
  {\doibase 10.1103/PhysRevD.93.054045} {\bibfield  {journal} {\bibinfo
  {journal} {Phys. Rev. D}\ }\textbf {\bibinfo {volume} {93}},\ \bibinfo
  {pages} {054045} (\bibinfo {year} {2016})},\ \Eprint
  {http://arxiv.org/abs/1511.04660} {arXiv:1511.04660 [hep-ph]} \BibitemShut
  {NoStop}%
\bibitem [{\citenamefont {Haque}\ and\ \citenamefont
  {Strickland}(2021)}]{Haque:2020eyj}%
  \BibitemOpen
  \bibfield  {author} {\bibinfo {author} {\bibfnamefont {N.}~\bibnamefont
  {Haque}}\ and\ \bibinfo {author} {\bibfnamefont {M.}~\bibnamefont
  {Strickland}},\ }\href {\doibase 10.1103/PhysRevC.103.L031901} {\bibfield
  {journal} {\bibinfo  {journal} {Phys. Rev. C}\ }\textbf {\bibinfo {volume}
  {103}},\ \bibinfo {pages} {031901} (\bibinfo {year} {2021})},\ \Eprint
  {http://arxiv.org/abs/2011.06938} {arXiv:2011.06938 [hep-ph]} \BibitemShut
  {NoStop}%
\bibitem [{\citenamefont {Huot}\ \emph {et~al.}(2007)\citenamefont {Huot},
  \citenamefont {Jeon},\ and\ \citenamefont {Moore}}]{Huot:2006ys}%
  \BibitemOpen
  \bibfield  {author} {\bibinfo {author} {\bibfnamefont {S.~C.}\ \bibnamefont
  {Huot}}, \bibinfo {author} {\bibfnamefont {S.}~\bibnamefont {Jeon}}, \ and\
  \bibinfo {author} {\bibfnamefont {G.~D.}\ \bibnamefont {Moore}},\ }\href
  {\doibase 10.1103/PhysRevLett.98.172303} {\bibfield  {journal} {\bibinfo
  {journal} {Phys. Rev. Lett.}\ }\textbf {\bibinfo {volume} {98}},\ \bibinfo
  {pages} {172303} (\bibinfo {year} {2007})},\ \Eprint
  {http://arxiv.org/abs/hep-ph/0608062} {arXiv:hep-ph/0608062} \BibitemShut
  {NoStop}%
\bibitem [{\citenamefont {Czajka}\ and\ \citenamefont
  {Mr\'owczy\'nski}(2014)}]{Czajka:2013lla}%
  \BibitemOpen
  \bibfield  {author} {\bibinfo {author} {\bibfnamefont {A.}~\bibnamefont
  {Czajka}}\ and\ \bibinfo {author} {\bibfnamefont {S.}~\bibnamefont
  {Mr\'owczy\'nski}},\ }\href {\doibase 10.5506/APhysPolBSupp.7.145} {\bibfield
   {journal} {\bibinfo  {journal} {Acta Phys. Polon. Supp.}\ }\textbf {\bibinfo
  {volume} {7}},\ \bibinfo {pages} {145} (\bibinfo {year} {2014})},\ \Eprint
  {http://arxiv.org/abs/1310.4709} {arXiv:1310.4709 [hep-th]} \BibitemShut
  {NoStop}%
\bibitem [{\citenamefont {Fotopoulos}\ and\ \citenamefont
  {Taylor}(1999)}]{Fotopoulos:1998es}%
  \BibitemOpen
  \bibfield  {author} {\bibinfo {author} {\bibfnamefont {A.}~\bibnamefont
  {Fotopoulos}}\ and\ \bibinfo {author} {\bibfnamefont {T.~R.}\ \bibnamefont
  {Taylor}},\ }\href {\doibase 10.1103/PhysRevD.59.061701} {\bibfield
  {journal} {\bibinfo  {journal} {Phys. Rev.}\ }\textbf {\bibinfo {volume}
  {D59}},\ \bibinfo {pages} {061701} (\bibinfo {year} {1999})},\ \Eprint
  {http://arxiv.org/abs/hep-th/9811224} {arXiv:hep-th/9811224 [hep-th]}
  \BibitemShut {NoStop}%
\bibitem [{\citenamefont {Kim}\ and\ \citenamefont {Rey}(2000)}]{Kim:1999sg}%
  \BibitemOpen
  \bibfield  {author} {\bibinfo {author} {\bibfnamefont {C.-j.}\ \bibnamefont
  {Kim}}\ and\ \bibinfo {author} {\bibfnamefont {S.-J.}\ \bibnamefont {Rey}},\
  }\href {\doibase 10.1016/S0550-3213(99)00532-5} {\bibfield  {journal}
  {\bibinfo  {journal} {Nucl. Phys.}\ }\textbf {\bibinfo {volume} {B564}},\
  \bibinfo {pages} {430} (\bibinfo {year} {2000})},\ \Eprint
  {http://arxiv.org/abs/hep-th/9905205} {arXiv:hep-th/9905205 [hep-th]}
  \BibitemShut {NoStop}%
\bibitem [{\citenamefont {Vazquez-Mozo}(1999)}]{VazquezMozo:1999ic}%
  \BibitemOpen
  \bibfield  {author} {\bibinfo {author} {\bibfnamefont {M.~A.}\ \bibnamefont
  {Vazquez-Mozo}},\ }\href {\doibase 10.1103/PhysRevD.60.106010} {\bibfield
  {journal} {\bibinfo  {journal} {Phys. Rev. D}\ }\textbf {\bibinfo {volume}
  {60}},\ \bibinfo {pages} {106010} (\bibinfo {year} {1999})},\ \Eprint
  {http://arxiv.org/abs/hep-th/9905030} {arXiv:hep-th/9905030} \BibitemShut
  {NoStop}%
\bibitem [{\citenamefont {Nieto}\ and\ \citenamefont
  {Tytgat}(1999)}]{Nieto:1999kc}%
  \BibitemOpen
  \bibfield  {author} {\bibinfo {author} {\bibfnamefont {A.}~\bibnamefont
  {Nieto}}\ and\ \bibinfo {author} {\bibfnamefont {M.~H.~G.}\ \bibnamefont
  {Tytgat}},\ }\href@noop {} {\  (\bibinfo {year} {1999})},\ \Eprint
  {http://arxiv.org/abs/hep-th/9906147} {arXiv:hep-th/9906147} \BibitemShut
  {NoStop}%
\bibitem [{\citenamefont {Du}\ \emph {et~al.}(2021)\citenamefont {Du},
  \citenamefont {Strickland},\ and\ \citenamefont {Tantary}}]{Du:2021jai}%
  \BibitemOpen
  \bibfield  {author} {\bibinfo {author} {\bibfnamefont {Q.}~\bibnamefont
  {Du}}, \bibinfo {author} {\bibfnamefont {M.}~\bibnamefont {Strickland}}, \
  and\ \bibinfo {author} {\bibfnamefont {U.}~\bibnamefont {Tantary}},\ }\href
  {\doibase 10.1007/JHEP08(2021)064} {\bibfield  {journal} {\bibinfo  {journal}
  {JHEP}\ }\textbf {\bibinfo {volume} {21}},\ \bibinfo {pages} {064} (\bibinfo
  {year} {2021})},\ \Eprint {http://arxiv.org/abs/2105.02101} {arXiv:2105.02101
  [hep-th]} \BibitemShut {NoStop}%
\bibitem [{\citenamefont {Andersen}\ \emph {et~al.}(2021)\citenamefont
  {Andersen}, \citenamefont {Du}, \citenamefont {Strickland},\ and\
  \citenamefont {Tantary}}]{Andersen:2021bgw}%
  \BibitemOpen
  \bibfield  {author} {\bibinfo {author} {\bibfnamefont {J.~O.}\ \bibnamefont
  {Andersen}}, \bibinfo {author} {\bibfnamefont {Q.}~\bibnamefont {Du}},
  \bibinfo {author} {\bibfnamefont {M.}~\bibnamefont {Strickland}}, \ and\
  \bibinfo {author} {\bibfnamefont {U.}~\bibnamefont {Tantary}},\ }\href@noop
  {} {\  (\bibinfo {year} {2021})},\ \Eprint {http://arxiv.org/abs/2111.12160}
  {arXiv:2111.12160 [hep-th]} \BibitemShut {NoStop}%
\bibitem [{\citenamefont {Luttinger}\ and\ \citenamefont
  {Ward}(1960)}]{Luttinger:1960ua}%
  \BibitemOpen
  \bibfield  {author} {\bibinfo {author} {\bibfnamefont {J.~M.}\ \bibnamefont
  {Luttinger}}\ and\ \bibinfo {author} {\bibfnamefont {J.~C.}\ \bibnamefont
  {Ward}},\ }\href {\doibase 10.1103/PhysRev.118.1417} {\bibfield  {journal}
  {\bibinfo  {journal} {Phys. Rev.}\ }\textbf {\bibinfo {volume} {118}},\
  \bibinfo {pages} {1417} (\bibinfo {year} {1960})}\BibitemShut {NoStop}%
\bibitem [{\citenamefont {Baym}(1962)}]{Baym:1962sx}%
  \BibitemOpen
  \bibfield  {author} {\bibinfo {author} {\bibfnamefont {G.}~\bibnamefont
  {Baym}},\ }\href {\doibase 10.1103/PhysRev.127.1391} {\bibfield  {journal}
  {\bibinfo  {journal} {Phys. Rev.}\ }\textbf {\bibinfo {volume} {127}},\
  \bibinfo {pages} {1391} (\bibinfo {year} {1962})}\BibitemShut {NoStop}%
\bibitem [{\citenamefont {Cornwall}\ \emph {et~al.}(1974)\citenamefont
  {Cornwall}, \citenamefont {Jackiw},\ and\ \citenamefont
  {Tomboulis}}]{Cornwall:1974vz}%
  \BibitemOpen
  \bibfield  {author} {\bibinfo {author} {\bibfnamefont {J.~M.}\ \bibnamefont
  {Cornwall}}, \bibinfo {author} {\bibfnamefont {R.}~\bibnamefont {Jackiw}}, \
  and\ \bibinfo {author} {\bibfnamefont {E.}~\bibnamefont {Tomboulis}},\ }\href
  {\doibase 10.1103/PhysRevD.10.2428} {\bibfield  {journal} {\bibinfo
  {journal} {Phys. Rev.}\ }\textbf {\bibinfo {volume} {D10}},\ \bibinfo {pages}
  {2428} (\bibinfo {year} {1974})}\BibitemShut {NoStop}%
\bibitem [{\citenamefont {Freedman}\ and\ \citenamefont
  {McLerran}(1977)}]{Freedman:1976ub}%
  \BibitemOpen
  \bibfield  {author} {\bibinfo {author} {\bibfnamefont {B.~A.}\ \bibnamefont
  {Freedman}}\ and\ \bibinfo {author} {\bibfnamefont {L.~D.}\ \bibnamefont
  {McLerran}},\ }\href {\doibase 10.1103/PhysRevD.16.1169} {\bibfield
  {journal} {\bibinfo  {journal} {Phys. Rev.}\ }\textbf {\bibinfo {volume}
  {D16}},\ \bibinfo {pages} {1169} (\bibinfo {year} {1977})}\BibitemShut
  {NoStop}%
\bibitem [{\citenamefont {Arrizabalaga}\ and\ \citenamefont
  {Smit}(2002)}]{Arrizabalaga:2002hn}%
  \BibitemOpen
  \bibfield  {author} {\bibinfo {author} {\bibfnamefont {A.}~\bibnamefont
  {Arrizabalaga}}\ and\ \bibinfo {author} {\bibfnamefont {J.}~\bibnamefont
  {Smit}},\ }\href {\doibase 10.1103/PhysRevD.66.065014} {\bibfield  {journal}
  {\bibinfo  {journal} {Phys. Rev.}\ }\textbf {\bibinfo {volume} {D66}},\
  \bibinfo {pages} {065014} (\bibinfo {year} {2002})},\ \Eprint
  {http://arxiv.org/abs/hep-ph/0207044} {arXiv:hep-ph/0207044 [hep-ph]}
  \BibitemShut {NoStop}%
\bibitem [{\citenamefont {Andersen}\ and\ \citenamefont
  {Strickland}(2005)}]{Andersen:2004re}%
  \BibitemOpen
  \bibfield  {author} {\bibinfo {author} {\bibfnamefont {J.~O.}\ \bibnamefont
  {Andersen}}\ and\ \bibinfo {author} {\bibfnamefont {M.}~\bibnamefont
  {Strickland}},\ }\href {\doibase 10.1103/PhysRevD.71.025011} {\bibfield
  {journal} {\bibinfo  {journal} {Phys. Rev.}\ }\textbf {\bibinfo {volume}
  {D71}},\ \bibinfo {pages} {025011} (\bibinfo {year} {2005})},\ \Eprint
  {http://arxiv.org/abs/hep-ph/0406163} {arXiv:hep-ph/0406163 [hep-ph]}
  \BibitemShut {NoStop}%
\bibitem [{\citenamefont {Andersen}\ \emph {et~al.}(2001)\citenamefont
  {Andersen}, \citenamefont {Braaten},\ and\ \citenamefont
  {Strickland}}]{Andersen:2000yj}%
  \BibitemOpen
  \bibfield  {author} {\bibinfo {author} {\bibfnamefont {J.~O.}\ \bibnamefont
  {Andersen}}, \bibinfo {author} {\bibfnamefont {E.}~\bibnamefont {Braaten}}, \
  and\ \bibinfo {author} {\bibfnamefont {M.}~\bibnamefont {Strickland}},\
  }\href {\doibase 10.1103/PhysRevD.63.105008} {\bibfield  {journal} {\bibinfo
  {journal} {Phys. Rev.}\ }\textbf {\bibinfo {volume} {D63}},\ \bibinfo {pages}
  {105008} (\bibinfo {year} {2001})},\ \Eprint
  {http://arxiv.org/abs/hep-ph/0007159} {arXiv:hep-ph/0007159 [hep-ph]}
  \BibitemShut {NoStop}%
\bibitem [{\citenamefont {Andersen}\ and\ \citenamefont
  {Strickland}(2001)}]{Andersen:2001ez}%
  \BibitemOpen
  \bibfield  {author} {\bibinfo {author} {\bibfnamefont {J.~O.}\ \bibnamefont
  {Andersen}}\ and\ \bibinfo {author} {\bibfnamefont {M.}~\bibnamefont
  {Strickland}},\ }\href {\doibase 10.1103/PhysRevD.64.105012} {\bibfield
  {journal} {\bibinfo  {journal} {Phys. Rev.}\ }\textbf {\bibinfo {volume}
  {D64}},\ \bibinfo {pages} {105012} (\bibinfo {year} {2001})},\ \Eprint
  {http://arxiv.org/abs/hep-ph/0105214} {arXiv:hep-ph/0105214 [hep-ph]}
  \BibitemShut {NoStop}%
\bibitem [{\citenamefont {Andersen}\ and\ \citenamefont
  {Kyllingstad}(2008)}]{Andersen:2008bz}%
  \BibitemOpen
  \bibfield  {author} {\bibinfo {author} {\bibfnamefont {J.~O.}\ \bibnamefont
  {Andersen}}\ and\ \bibinfo {author} {\bibfnamefont {L.}~\bibnamefont
  {Kyllingstad}},\ }\href {\doibase 10.1103/PhysRevD.78.076008} {\bibfield
  {journal} {\bibinfo  {journal} {Phys.\ Rev.\ D}\ }\textbf {\bibinfo {volume}
  {78}},\ \bibinfo {pages} {076008} (\bibinfo {year} {2008})},\ \Eprint
  {http://arxiv.org/abs/0805.4478} {arXiv:0805.4478 [hep-ph]} \BibitemShut
  {NoStop}%
\bibitem [{\citenamefont {Andersen}\ \emph {et~al.}(2009)\citenamefont
  {Andersen}, \citenamefont {Strickland},\ and\ \citenamefont
  {Su}}]{Andersen:2009tw}%
  \BibitemOpen
  \bibfield  {author} {\bibinfo {author} {\bibfnamefont {J.~O.}\ \bibnamefont
  {Andersen}}, \bibinfo {author} {\bibfnamefont {M.}~\bibnamefont
  {Strickland}}, \ and\ \bibinfo {author} {\bibfnamefont {N.}~\bibnamefont
  {Su}},\ }\href {\doibase 10.1103/PhysRevD.80.085015} {\bibfield  {journal}
  {\bibinfo  {journal} {Phys. Rev.}\ }\textbf {\bibinfo {volume} {D80}},\
  \bibinfo {pages} {085015} (\bibinfo {year} {2009})},\ \Eprint
  {http://arxiv.org/abs/0906.2936} {arXiv:0906.2936 [hep-ph]} \BibitemShut
  {NoStop}%
\bibitem [{\citenamefont {Du}\ \emph {et~al.}(2020)\citenamefont {Du},
  \citenamefont {Strickland}, \citenamefont {Tantary},\ and\ \citenamefont
  {Zhang}}]{Du:2020odw}%
  \BibitemOpen
  \bibfield  {author} {\bibinfo {author} {\bibfnamefont {Q.}~\bibnamefont
  {Du}}, \bibinfo {author} {\bibfnamefont {M.}~\bibnamefont {Strickland}},
  \bibinfo {author} {\bibfnamefont {U.}~\bibnamefont {Tantary}}, \ and\
  \bibinfo {author} {\bibfnamefont {B.-W.}\ \bibnamefont {Zhang}},\ }\href
  {\doibase 10.1007/JHEP09(2020)038} {\bibfield  {journal} {\bibinfo  {journal}
  {JHEP}\ }\textbf {\bibinfo {volume} {09}},\ \bibinfo {pages} {038} (\bibinfo
  {year} {2020})},\ \Eprint {http://arxiv.org/abs/2006.02617} {arXiv:2006.02617
  [hep-ph]} \BibitemShut {NoStop}%
\bibitem [{\citenamefont {Ashmore}(1972)}]{Ashmore:1972uj}%
  \BibitemOpen
  \bibfield  {author} {\bibinfo {author} {\bibfnamefont {J.~F.}\ \bibnamefont
  {Ashmore}},\ }\href {\doibase 10.1007/BF02824407} {\bibfield  {journal}
  {\bibinfo  {journal} {Lett. Nuovo Cim.}\ }\textbf {\bibinfo {volume} {4}},\
  \bibinfo {pages} {289} (\bibinfo {year} {1972})}\BibitemShut {NoStop}%
\bibitem [{\citenamefont {Bollini}\ and\ \citenamefont
  {Giambiagi}(1972)}]{Bollini:1972ui}%
  \BibitemOpen
  \bibfield  {author} {\bibinfo {author} {\bibfnamefont {C.~G.}\ \bibnamefont
  {Bollini}}\ and\ \bibinfo {author} {\bibfnamefont {J.~J.}\ \bibnamefont
  {Giambiagi}},\ }\href {\doibase 10.1007/BF02895558} {\bibfield  {journal}
  {\bibinfo  {journal} {Nuovo Cim. B}\ }\textbf {\bibinfo {volume} {12}},\
  \bibinfo {pages} {20} (\bibinfo {year} {1972})}\BibitemShut {NoStop}%
\bibitem [{\citenamefont {{'t Hooft}}\ and\ \citenamefont
  {Veltman}(1972)}]{THOOFT1972189}%
  \BibitemOpen
  \bibfield  {author} {\bibinfo {author} {\bibfnamefont {G.}~\bibnamefont {{'t
  Hooft}}}\ and\ \bibinfo {author} {\bibfnamefont {M.}~\bibnamefont
  {Veltman}},\ }\href {\doibase https://doi.org/10.1016/0550-3213(72)90279-9}
  {\bibfield  {journal} {\bibinfo  {journal} {Nuclear Physics B}\ }\textbf
  {\bibinfo {volume} {44}},\ \bibinfo {pages} {189} (\bibinfo {year}
  {1972})}\BibitemShut {NoStop}%
\bibitem [{\citenamefont {Siegel}(1979)}]{Siegel:1979wq}%
  \BibitemOpen
  \bibfield  {author} {\bibinfo {author} {\bibfnamefont {W.}~\bibnamefont
  {Siegel}},\ }\href {\doibase 10.1016/0370-2693(79)90282-X} {\bibfield
  {journal} {\bibinfo  {journal} {Phys. Lett. B}\ }\textbf {\bibinfo {volume}
  {84}},\ \bibinfo {pages} {193} (\bibinfo {year} {1979})}\BibitemShut
  {NoStop}%
\bibitem [{\citenamefont {Brink}\ \emph {et~al.}(1977)\citenamefont {Brink},
  \citenamefont {Schwarz},\ and\ \citenamefont {Scherk}}]{Brink:1976bc}%
  \BibitemOpen
  \bibfield  {author} {\bibinfo {author} {\bibfnamefont {L.}~\bibnamefont
  {Brink}}, \bibinfo {author} {\bibfnamefont {J.~H.}\ \bibnamefont {Schwarz}},
  \ and\ \bibinfo {author} {\bibfnamefont {J.}~\bibnamefont {Scherk}},\ }\href
  {\doibase 10.1016/0550-3213(77)90328-5} {\bibfield  {journal} {\bibinfo
  {journal} {Nucl. Phys. B}\ }\textbf {\bibinfo {volume} {121}},\ \bibinfo
  {pages} {77} (\bibinfo {year} {1977})}\BibitemShut {NoStop}%
\bibitem [{\citenamefont {Gliozzi}\ \emph {et~al.}(1977)\citenamefont
  {Gliozzi}, \citenamefont {Scherk},\ and\ \citenamefont
  {Olive}}]{Gliozzi:1976qd}%
  \BibitemOpen
  \bibfield  {author} {\bibinfo {author} {\bibfnamefont {F.}~\bibnamefont
  {Gliozzi}}, \bibinfo {author} {\bibfnamefont {J.}~\bibnamefont {Scherk}}, \
  and\ \bibinfo {author} {\bibfnamefont {D.~I.}\ \bibnamefont {Olive}},\ }\href
  {\doibase 10.1016/0550-3213(77)90206-1} {\bibfield  {journal} {\bibinfo
  {journal} {Nucl. Phys. B}\ }\textbf {\bibinfo {volume} {122}},\ \bibinfo
  {pages} {253} (\bibinfo {year} {1977})}\BibitemShut {NoStop}%
\bibitem [{\citenamefont {Avdeev}\ and\ \citenamefont
  {Vladimirov}(1983)}]{Avdeev:1982xy}%
  \BibitemOpen
  \bibfield  {author} {\bibinfo {author} {\bibfnamefont {L.~V.}\ \bibnamefont
  {Avdeev}}\ and\ \bibinfo {author} {\bibfnamefont {A.~A.}\ \bibnamefont
  {Vladimirov}},\ }\href {\doibase 10.1016/0550-3213(83)90437-6} {\bibfield
  {journal} {\bibinfo  {journal} {Nucl. Phys. B}\ }\textbf {\bibinfo {volume}
  {219}},\ \bibinfo {pages} {262} (\bibinfo {year} {1983})}\BibitemShut
  {NoStop}%
\bibitem [{\citenamefont {Capper}\ \emph {et~al.}(1980)\citenamefont {Capper},
  \citenamefont {Jones},\ and\ \citenamefont {{Van
  Nieuwenhuizen}}}]{CAPPER1980479}%
  \BibitemOpen
  \bibfield  {author} {\bibinfo {author} {\bibfnamefont {D.}~\bibnamefont
  {Capper}}, \bibinfo {author} {\bibfnamefont {D.}~\bibnamefont {Jones}}, \
  and\ \bibinfo {author} {\bibfnamefont {P.}~\bibnamefont {{Van
  Nieuwenhuizen}}},\ }\href {\doibase
  https://doi.org/10.1016/0550-3213(80)90244-8} {\bibfield  {journal} {\bibinfo
   {journal} {Nuclear Physics B}\ }\textbf {\bibinfo {volume} {167}},\ \bibinfo
  {pages} {479} (\bibinfo {year} {1980})}\BibitemShut {NoStop}%
\bibitem [{\citenamefont {Bender}\ and\ \citenamefont
  {Wu}(1969)}]{Bender:1969si}%
  \BibitemOpen
  \bibfield  {author} {\bibinfo {author} {\bibfnamefont {C.~M.}\ \bibnamefont
  {Bender}}\ and\ \bibinfo {author} {\bibfnamefont {T.~T.}\ \bibnamefont
  {Wu}},\ }\href {\doibase 10.1103/PhysRev.184.1231} {\bibfield  {journal}
  {\bibinfo  {journal} {Phys. Rev.}\ }\textbf {\bibinfo {volume} {184}},\
  \bibinfo {pages} {1231} (\bibinfo {year} {1969})}\BibitemShut {NoStop}%
\bibitem [{\citenamefont {Bender}\ and\ \citenamefont
  {Wu}(1973)}]{Bender:1973rz}%
  \BibitemOpen
  \bibfield  {author} {\bibinfo {author} {\bibfnamefont {C.~M.}\ \bibnamefont
  {Bender}}\ and\ \bibinfo {author} {\bibfnamefont {T.~T.}\ \bibnamefont
  {Wu}},\ }\href {\doibase 10.1103/PhysRevD.7.1620} {\bibfield  {journal}
  {\bibinfo  {journal} {Phys. Rev. D}\ }\textbf {\bibinfo {volume} {7}},\
  \bibinfo {pages} {1620} (\bibinfo {year} {1973})}\BibitemShut {NoStop}%
\bibitem [{\citenamefont {Janke}\ and\ \citenamefont
  {Kleinert}(1995)}]{Janke:1995zz}%
  \BibitemOpen
  \bibfield  {author} {\bibinfo {author} {\bibfnamefont {W.}~\bibnamefont
  {Janke}}\ and\ \bibinfo {author} {\bibfnamefont {H.}~\bibnamefont
  {Kleinert}},\ }\href {\doibase 10.1103/PhysRevLett.75.2787} {\bibfield
  {journal} {\bibinfo  {journal} {Phys. Rev. Lett.}\ }\textbf {\bibinfo
  {volume} {75}},\ \bibinfo {pages} {2787} (\bibinfo {year} {1995})},\ \Eprint
  {http://arxiv.org/abs/quant-ph/9502019} {arXiv:quant-ph/9502019} \BibitemShut
  {NoStop}%
\bibitem [{\citenamefont {Kleinert}\ and\ \citenamefont
  {Janke}(1995)}]{Kleinert:1995hc}%
  \BibitemOpen
  \bibfield  {author} {\bibinfo {author} {\bibfnamefont {H.}~\bibnamefont
  {Kleinert}}\ and\ \bibinfo {author} {\bibfnamefont {W.}~\bibnamefont
  {Janke}},\ }\href {\doibase 10.1016/0375-9601(95)00521-4} {\bibfield
  {journal} {\bibinfo  {journal} {Phys. Lett. A}\ }\textbf {\bibinfo {volume}
  {206}},\ \bibinfo {pages} {283} (\bibinfo {year} {1995})},\ \Eprint
  {http://arxiv.org/abs/quant-ph/9509005} {arXiv:quant-ph/9509005} \BibitemShut
  {NoStop}%
\bibitem [{\citenamefont {Bardeen}\ \emph {et~al.}(1978)\citenamefont
  {Bardeen}, \citenamefont {Buras}, \citenamefont {Duke},\ and\ \citenamefont
  {Muta}}]{Bardeen:1978yd}%
  \BibitemOpen
  \bibfield  {author} {\bibinfo {author} {\bibfnamefont {W.~A.}\ \bibnamefont
  {Bardeen}}, \bibinfo {author} {\bibfnamefont {A.~J.}\ \bibnamefont {Buras}},
  \bibinfo {author} {\bibfnamefont {D.~W.}\ \bibnamefont {Duke}}, \ and\
  \bibinfo {author} {\bibfnamefont {T.}~\bibnamefont {Muta}},\ }\href {\doibase
  10.1103/PhysRevD.18.3998} {\bibfield  {journal} {\bibinfo  {journal} {Phys.
  Rev. D}\ }\textbf {\bibinfo {volume} {18}},\ \bibinfo {pages} {3998}
  (\bibinfo {year} {1978})}\BibitemShut {NoStop}%
\bibitem [{\citenamefont {'t~Hooft}(1973)}]{tHooft:1973mfk}%
  \BibitemOpen
  \bibfield  {author} {\bibinfo {author} {\bibfnamefont {G.}~\bibnamefont
  {'t~Hooft}},\ }\href {\doibase 10.1016/0550-3213(73)90376-3} {\bibfield
  {journal} {\bibinfo  {journal} {Nucl. Phys. B}\ }\textbf {\bibinfo {volume}
  {61}},\ \bibinfo {pages} {455} (\bibinfo {year} {1973})}\BibitemShut
  {NoStop}%
\bibitem [{\citenamefont {Weinberg}(1973)}]{Weinberg:1973xwm}%
  \BibitemOpen
  \bibfield  {author} {\bibinfo {author} {\bibfnamefont {S.}~\bibnamefont
  {Weinberg}},\ }\href {\doibase 10.1103/PhysRevD.8.3497} {\bibfield  {journal}
  {\bibinfo  {journal} {Phys. Rev. D}\ }\textbf {\bibinfo {volume} {8}},\
  \bibinfo {pages} {3497} (\bibinfo {year} {1973})}\BibitemShut {NoStop}%
\bibitem [{\citenamefont {Stevenson}(1982)}]{Stevenson:1982wn}%
  \BibitemOpen
  \bibfield  {author} {\bibinfo {author} {\bibfnamefont {P.}~\bibnamefont
  {Stevenson}},\ }\href {\doibase 10.1016/0550-3213(82)90325-X} {\bibfield
  {journal} {\bibinfo  {journal} {Nucl. Phys. B}\ }\textbf {\bibinfo {volume}
  {203}},\ \bibinfo {pages} {472} (\bibinfo {year} {1982})}\BibitemShut
  {NoStop}%
\bibitem [{\citenamefont {Lepage}(1989)}]{Lepage:1989hf}%
  \BibitemOpen
  \bibfield  {author} {\bibinfo {author} {\bibfnamefont {G.~P.}\ \bibnamefont
  {Lepage}},\ }in\ \href@noop {} {\emph {\bibinfo {booktitle} {{Theoretical
  Advanced Study Institute in Elementary Particle Physics}}}}\ (\bibinfo {year}
  {1989})\ \Eprint {http://arxiv.org/abs/hep-ph/0506330} {arXiv:hep-ph/0506330}
  \BibitemShut {NoStop}%
\bibitem [{\citenamefont {Quevedo}\ \emph {et~al.}(2010)\citenamefont
  {Quevedo}, \citenamefont {Krippendorf},\ and\ \citenamefont
  {Schlotterer}}]{Quevedo:2010ui}%
  \BibitemOpen
  \bibfield  {author} {\bibinfo {author} {\bibfnamefont {F.}~\bibnamefont
  {Quevedo}}, \bibinfo {author} {\bibfnamefont {S.}~\bibnamefont
  {Krippendorf}}, \ and\ \bibinfo {author} {\bibfnamefont {O.}~\bibnamefont
  {Schlotterer}},\ }\href@noop {} {\  (\bibinfo {year} {2010})},\ \Eprint
  {http://arxiv.org/abs/1011.1491} {arXiv:1011.1491 [hep-th]} \BibitemShut
  {NoStop}%
\bibitem [{\citenamefont {Bertolini}(2015)}]{bertolini2015lectures}%
  \BibitemOpen
  \bibfield  {author} {\bibinfo {author} {\bibfnamefont {M.}~\bibnamefont
  {Bertolini}},\ }\href@noop {} {\bibfield  {journal} {\bibinfo  {journal}
  {SISSA -- International School for Advanced Studies}\ } (\bibinfo {year}
  {2015})},\ \Eprint
  {http://arxiv.org/abs/https://people.sissa.it/$\sim$bertmat/susycourse.pdf}
  {https://people.sissa.it/$\sim$bertmat/susycourse.pdf} \BibitemShut {NoStop}%
\bibitem [{\citenamefont {Yamada}\ and\ \citenamefont
  {Yaffe}(2006)}]{Yamada:2006rx}%
  \BibitemOpen
  \bibfield  {author} {\bibinfo {author} {\bibfnamefont {D.}~\bibnamefont
  {Yamada}}\ and\ \bibinfo {author} {\bibfnamefont {L.~G.}\ \bibnamefont
  {Yaffe}},\ }\href {\doibase 10.1088/1126-6708/2006/09/027} {\bibfield
  {journal} {\bibinfo  {journal} {JHEP}\ }\textbf {\bibinfo {volume} {09}},\
  \bibinfo {pages} {027} (\bibinfo {year} {2006})},\ \Eprint
  {http://arxiv.org/abs/hep-th/0602074} {arXiv:hep-th/0602074 [hep-th]}
  \BibitemShut {NoStop}%
\bibitem [{\citenamefont {D'Hoker}\ and\ \citenamefont
  {Phong}(1999)}]{DHoker:1999yni}%
  \BibitemOpen
  \bibfield  {author} {\bibinfo {author} {\bibfnamefont {E.}~\bibnamefont
  {D'Hoker}}\ and\ \bibinfo {author} {\bibfnamefont {D.~H.}\ \bibnamefont
  {Phong}},\ }in\ \href@noop {} {\emph {\bibinfo {booktitle} {{Theoretical
  physics at the end of the twentieth century. Proceedings, Summer School,
  Banff, Canada, June 27-July 10, 1999}}}}\ (\bibinfo {year} {1999})\ pp.\
  \bibinfo {pages} {1--125},\ \Eprint {http://arxiv.org/abs/hep-th/9912271}
  {arXiv:hep-th/9912271 [hep-th]} \BibitemShut {NoStop}%
\bibitem [{\citenamefont {Kovacs}(1999)}]{Kovacs:1999fx}%
  \BibitemOpen
  \bibfield  {author} {\bibinfo {author} {\bibfnamefont {S.}~\bibnamefont
  {Kovacs}},\ }\emph {\bibinfo {title} {{N=4 supersymmetric Yang-Mills theory
  and the AdS / SCFT correspondence}}},\ \href@noop {} {Ph.D. thesis},\
  \bibinfo  {school} {Rome U., Tor Vergata} (\bibinfo {year} {1999}),\ \Eprint
  {http://arxiv.org/abs/hep-th/9908171} {arXiv:hep-th/9908171 [hep-th]}
  \BibitemShut {NoStop}%
\bibitem [{\citenamefont {Blaizot}\ \emph {et~al.}(2007)\citenamefont
  {Blaizot}, \citenamefont {Iancu}, \citenamefont {Kraemmer},\ and\
  \citenamefont {Rebhan}}]{Blaizot:2006tk}%
  \BibitemOpen
  \bibfield  {author} {\bibinfo {author} {\bibfnamefont {J.~P.}\ \bibnamefont
  {Blaizot}}, \bibinfo {author} {\bibfnamefont {E.}~\bibnamefont {Iancu}},
  \bibinfo {author} {\bibfnamefont {U.}~\bibnamefont {Kraemmer}}, \ and\
  \bibinfo {author} {\bibfnamefont {A.}~\bibnamefont {Rebhan}},\ }\href
  {\doibase 10.1088/1126-6708/2007/06/035} {\bibfield  {journal} {\bibinfo
  {journal} {JHEP}\ }\textbf {\bibinfo {volume} {06}},\ \bibinfo {pages} {035}
  (\bibinfo {year} {2007})},\ \Eprint {http://arxiv.org/abs/hep-ph/0611393}
  {arXiv:hep-ph/0611393 [hep-ph]} \BibitemShut {NoStop}%
\end{thebibliography}%

\end{document}